\documentclass{aa}
\usepackage[varg]{txfonts}
\usepackage{natbib}
\bibpunct{(}{)}{;}{a}{}{,} 

\usepackage{amsmath}
\usepackage{graphicx}
\usepackage{graphics}
\usepackage{epsfig}
\usepackage{pdfpages} 
\usepackage{blindtext}
\usepackage{supertabular}
\usepackage{afterpage}
\usepackage{lscape}
\usepackage{tablefootnote}
\usepackage{subcaption}
\usepackage[flushleft]{threeparttable}
\usepackage{color, colortbl}
\usepackage[normalem]{ulem}
\usepackage{multirow}
\definecolor{lightgray}{gray}{0.9}
\usepackage{chemformula}

\usepackage{txfonts}

\usepackage{natbib,twoopt}
\usepackage[colorlinks=true,linkcolor=blue,citecolor=blue,urlcolor=magenta,breaklinks=true]{hyperref} 
\bibpunct{(}{)}{;}{a}{}{,}             
\makeatletter
  \newcommandtwoopt{\citeads}[3][][]{\href{http://adsabs.harvard.edu/abs/#3}%
    {\def\hyper@linkstart##1##2{}%
     \let\hyper@linkend\@empty\citealp[#1][#2]{#3}}}
  \newcommandtwoopt{\citepads}[3][][]{\href{http://adsabs.harvard.edu/abs/#3}%
    {\def\hyper@linkstart##1##2{}%
     \let\hyper@linkend\@empty\citep[#1][#2]{#3}}}
  \newcommandtwoopt{\citetads}[3][][]{\href{http://adsabs.harvard.edu/abs/#3}%
    {\def\hyper@linkstart##1##2{}%
     \let\hyper@linkend\@empty\citet[#1][#2]{#3}}}
  \newcommandtwoopt{\citeyearads}[3][][]%
    {\href{http://adsabs.harvard.edu/abs/#3}
    {\def\hyper@linkstart##1##2{}%
     \let\hyper@linkend\@empty\citeyear[#1][#2]{#3}}}
\makeatother

\begin{document}
\title{DUNE: Dust depletion UNified method across cosmic time and Environments} 

\author{Christina~Konstantopoulou\inst{\ref{inst-geneva}}\thanks{\email{christina.konstantopoulou@unige.ch}},
        Annalisa~De~Cia\inst{\ref{inst-geneva},\ref{inst-esogarch}}
        Jens-Kristian~Krogager\inst{\ref{inst-lyon}, \ref{inst-chile}},     
        C\'{e}dric~Ledoux\inst{\ref{inst-eso}},
        Julia~Roman-Duval\inst{\ref{inst-stsc}},        Edward~B.~Jenkins\inst{\ref{inst-princeton}},
        Tanita~Ramburuth-Hurt\inst{\ref{inst-geneva}},
        and Anna~Velichko\inst{\ref{inst-geneva},\ref{inst-ukraine}}}


\institute{Department of Astronomy, University of Geneva, Chemin Pegasi 51, 1290 Versoix, Switzerland \label{inst-geneva}
\and{European Southern Observatory, Karl-Schwarzschild Str. 2, 85748 Garching bei München, Germany\label{inst-esogarch}
\and Centre de Recherche Astrophysique de Lyon, Univ. Claude Bernard Lyon 1, 9 Av. Charles Andr\'{e}, 69230 Saint-Genis-Laval, France\label{inst-lyon}
\and French-Chilean Laboratory for Astronomy (FCLA), CNRS-IRL3386, U. de Chile, Camino el Observatorio 1515, Casilla 36-D, Santiago, Chile  \label{inst-chile}
\and European Southern Observatory, Alonso de C\'{o}rdova 3107, Vitacura, Casilla 19001, Santiago, Chile \label{inst-eso}
\and Space Telescope Science Institute, 3700 San Martin Drive, Baltimore, MD21218, USA \label{inst-stsc}
\and Dept. of Astrophysical Sciences, Princeton University, Princeton, NJ 08544, USA \label{inst-princeton}$^{\dagger}$
\and Institute of Astronomy, Kharkiv National University, 4 Svobody Sq., Kharkiv, 61022, Ukraine
\label{inst-ukraine}}}

\date{Received xxx; Accepted xxx}

 \abstract{We present a novel method to characterize dust depletion, namely, the depletion of metals into dust grains. We used observed correlations among relative abundances combining a total of 17 metals in diverse galactic environments, including the Milky Way (MW),  Large Magellanic Cloud (LMC),  Small Magellanic Cloud (SMC), and damped Lyman-$\alpha$ absorbers (DLAs) towards quasars and gamma-ray bursts (GRBs). We only considered the relative abundances of metals that qualify as tracers of dust and we used all available dust tracers. We find linear correlations among all studied dust tracers in a multidimensional space, where each dimension corresponds to an individual dust tracer. The fit to the linear correlations among the dust tracers describes the tendencies of different elements when depleting into dust grains. We determined the overall strength of dust depletion, $\Delta$, along individual lines of sight, based on the correlations among different dust tracers. We avoided any preference for 
 specific dust tracers or any other assumptions by including all available dust tracers in this multidimensional space. We also determined the dust depletion of Kr, C, O, Cl, P, Zn, Ge, Mg, Cu, Si, Fe, Ni, and Ti.
 Finally, we offer simple guidelines for the application of the method to the study of the observed patterns of abundances and relative abundances. This has allowed for a straightforward determination of the overall strength of depletion and the dust depletion of individual elements. We also obtained an estimate for the gas-phase metallicity and identified any additional deviations due to the nucleosynthesis of specific stellar populations. Thus, we have established a unified methodology for characterizing dust depletion across cosmic time and diverse galactic environments, offering a valuable new approach to the study of dust depletion in studies of the chemical evolution of galaxies.  
 }

\keywords{galaxies: evolution - ISM: dust, extinction}
\authorrunning{C. Konstantopoulou et al.}
\titlerunning{DUNE}
\maketitle

\let\thefootnote\relax\footnotetext{$^{\dagger}$ Deceased.}

\section{Introduction}

Metals are produced within stars and released into the interstellar medium (ISM) of galaxies. While some of these metals remain in the gas-phase of the ISM, a significant amount is instead incorporated into dust grains, an effect called dust depletion \citep{Field1974, Savage1996, Phillips1982, Savaglio2003, Jenkins2009, DeCia2016, Roman-Duval2021, Konstantopoulou2022}. Gas-phase metal abundances, resulting from column density measurements of metals, \ion{H}{i} and \ion{H}{ii}, obtained through absorption-line spectroscopy, can vary depending on dust depletion, but also on metallicity variations in the gas and the nucleosynthetic history of metals. Understanding dust depletion is still a challenge, especially because of the difficulties in disentangling these different phenomena that affect the observed abundances. 

Dust depletion is a critical phenomenon within the ISM of galaxies \citep{Roman-Duval2022a, Konstantopoulou2024, DeCia2024}. Understanding dust depletion is important for several reasons. Applying depletion corrections for damped Lyman-$\alpha$ absorbers (DLAs) is essential for tracking the chemical enrichment of the universe over cosmic time and it provides an estimate of the dust-to-gas ratio that is independent of FIR data, which can suffer from limited sensitivity and resolution \citep{Roman-Duval2014, Roman-Duval2021}. Therefore, depletion corrections improve the accuracy of dust and gas mass estimates across all redshifts and help us understand the variations of dust properties in galaxies. Therefore, a comprehensive study of dust depletion is crucial to understanding how dust is formed and how metals are distributed within the ISM of galaxies. During dust formation, metals are collected from the gas and incorporated into dust grains, while dust destruction (e.g., by SN shocks) releases them back into the gas-phase of the ISM. The interplay of metals between gas-phase, dust-phase, and stars is fundamental for the chemical evolution of galaxies.

\citet{Jenkins2009} developed a unified scheme of elemental depletions in the ISM of the Milky Way. This method arises from the discovery that the observed abundances [X/H] of different metals are correlated with each other. For example, [Mg/H] is correlated with [Fe/H], which, in turn, is correlated with [P/H] and so on \citep[see Fig. 1 of][]{Jenkins2009}. The observed abundances are used as a proxy for the elemental depletions. This approach shows that different lines of sight have a continuous range of depletions (as opposed to discrete depletion patterns \citep[e.g.,][]{Savage1996}. The overall strength of dust depletion along an individual line of sight is represented by the factor F$_{\ast}$ in \citet{Jenkins2009}. The value of F$_{\ast}$ is defined to be F$_{\ast}$ = 1, for lines of sight exhibiting the strongest depletions, while for those with the lowest depletions F$_{\ast}$ is assigned to be F$_{\ast}$ = 0. This scale is somewhat arbitrary because the F$_{\ast}$ = 0 and F$_{\ast}$ = 1 end points are tied to the particular sample studied in \citet{Jenkins2009}. In the \citet{Jenkins2009} formalism, the depletion of individual elements increases with F$_{\ast}$. This increase in depletion is often associated with denser and/or colder gas environments. Lines of sight with lower depletion may indicate that grains have undergone partial destruction through sputtering in interstellar shocks and/or they have not grown as much as the grains in the denser ISM from the accretion of metals from the gas-phase. This progression from low-density diffuse atomic gas to higher-density diffuse molecular clouds can be interpreted as grain growth in the ISM. Hereafter, we refer this method, developed by \citet{Jenkins2009}, as the "F$_{\ast}$ method". 

The main strength of the F$_{\ast}$ method is that it uses all the observed abundances of metals, without favoring any metal in particular, to estimate the overall strength of depletion along individual lines of sight. It identifies correlations among different elements and finds continuous trends in elemental depletions, characterizing each line of sight with a single measure of the overall strength of depletion. However, it assumes that the total (gas+dust) abundance in the ISM is equal to the photospheric abundances of young stars. Moreover, the usage of all the observed abundances of metals is only valid for the \citet{Jenkins2009} sample. For other samples in other galaxies or even the MW, we have to use the Fe depletion to define F$_{\ast}$ based on the \citet{Jenkins2009} coefficients.
In addition, the transport of metals from stars to the ISM is unaccounted for. More specifically, the determination of abundances of metals is complicated by variations in metallicity and nucleosynthesis effects, such as $\alpha$-element enhancement, in the gas of the ISM. These effects are not trivial to disentangle and can affect the observed trends, which are interpreted as dust depletion within the F$_{\ast}$ method. Moreover, the F$_{\ast}$ method assumes a fixed metallicity for the gas, and specifically for the MW it assumes solar metallicity. This may serve as a reasonable approximation for the MW but cannot be assumed for other galaxies. To use the F$_{\ast}$ method in other galaxies, we have to assume a different metallicity than solar. Additionally, the F$_{\ast}$ scale is tied to specific lines of sight in the \citet{Jenkins2009} sample and, thus, this scale is arbitrary. Therefore, we must use Fe depletions to define F$_{\ast}$, based on the \citet{Jenkins2009} coefficients.
In fact, even the metallicity of the MW remains puzzling. While \citet{Ritchey2023} have found that the metallicity of the MW is fairly homogeneous, large  variations in metallicity have been observed in the neutral ISM in the vicinity of the Sun \citep{DeCia2021}. One important difference between the two studies is that \citet{Ritchey2023} included  volatile elements in their analysis,  thus probing higher metallicity gas; whereas \citet{DeCia2021} excluded volatile elements, probing mostly lower metallicity gas. Hence, while the F$_{\ast}$ method offers valuable insights and has become one of the key pillars for the study of dust depletion, its assumptions should be approached with caution when we are considering  environments beyond the MW.

\citet{DeCia2016} developed an independent study of dust depletion, based on observations of the MW and distant galaxies. \citet{DeCia2018} extended this, to study dust depletion using observations in the Magellanic Clouds. This method uses relative abundances (meaning abundance ratios) of metals as dust tracers rather than observed abundances to estimate the depletion of different elements because different metals have different tendencies to deplete into dust grains. A relative abundance [X/Y] can be a dust tracer if X and Y follow each other in nucleosynthesis, but have very different refractory properties. \citet{DeCia2016} mainly used [Zn/Fe] as a dust tracer, but note that others are possible (e.g., [Si/Ti]). The depletion of element X is then obtained from the observed correlation between [X/Zn] (a quantity dominated by the depletion of X because Zn is rather volatile) and the dust tracer [Zn/Fe]. Hereafter, we refer to this method, developed by \citet{DeCia2016}, as the "relative method".

The main strength of the relative method is the use of relative abundances of metals, making it possible to disentangle effects, such as dust depletion, metallicity, and nucleosynthesis. If there are variations in metallicity, the relative abundances are not affected. On the other hand, variations in specific metals that do not follow dust depletion, such $\alpha$-element enhancements, can be identified and are indeed observed in the ISM \citep{DeCia2024, Velichko2024}. Consequently, this methodology can be applied across a wide range of galactic environments, including distant galaxies and low-metallicity systems, extending the study of dust depletion beyond the MW. However, the main weakness of this method is that Zn has a preferential role in the method's formalism. The use of [X/Zn] introduces a dependency on the depletion of Zn, which is built-in by assuming a certain slope for the expected depletion of Zn with the dust tracer [Zn/Fe]. The impact of this assumption on the recovered depletions in DLAs is investigated in \citet{Roman-Duval2022b}.
Nevertheless, the accuracy of the estimated Zn depletion is solely based on Zn abundances and may not be precise. Furthermore, there is an uncertainty regarding the reliability of [Zn/Fe] as a dust tracer. While [Zn/Fe] is often considered a good tracer of dust \citep[e.g.,][]{DeCia2016, DeCia2018, Konstantopoulou2022}, due to the fact that Zn and Fe trace each other in the Solar neighborhood ([Zn/Fe] $\sim$ 0 for $-2 <$ [Fe/H] $< 0$ dex), the nucleosynthetic behavior of Zn is complex as it does not strictly belong to the Fe-peak or $\alpha$-element group. This means that Zn can show an enhancement that resembles $\alpha$-element enhancement, albeit with a smaller amplitude \citep[e.g.,][]{Nissen2011, Barbuy2015, Duffau2017}. This effect has recently been observed in the ISM as well \citep{DeCia2024, Velichko2024}. Finally, the methodology of \citet{DeCia2016} also made an assumption on the shape of the $\alpha$-element enhancement curves to be able to convert the observed relative abundances [X/Zn] to depletions. Hence,  the relative method offers a new approach to studying dust depletion across diverse galactic environments and its reliance on Zn, as well as any assumptions placed on nucleosynthesis; however, it does introduce caveats that need to be taken into consideration.

Motivated by the above two existing methods for estimating dust depletion, the aim of this paper is to combine their strengths and overcome their limitations, by developing a Dust depletion UNified method across cosmic time and Environments (DUNE). In this method, we use correlations among all relative abundances that are valid dust tracers, to estimate the dust depletion of different metals in diverse galactic environments, including the MW, LMC, SMC, and QSO- and GRB-DLAs.

The paper is organized as follows. In Sect. \ref{sec:data}, we briefly present the data samples compiled from the literature. The development of the method is presented in Sect. \ref{sec:meth_res}. We discuss our results in Sect. \ref{sec: discussion}. Finally, we summarize and conclude in Sect. \ref{sec: conclusions}.
Throughout the paper, we use a linear unit for the column densities, N, in terms of ions cm$^{-2}$. We refer to relative abundances of elements X and Y as $[\rm X/Y] \equiv$ log$\frac{N(\rm{X})}{N(\rm Y)} - $ log$\frac{N(\rm X)_{\odot}}{N(\rm Y)_{\odot}}$, where reference solar abundances are taken from \citet{Asplund2021} following the recommendations of \citet{Lodders2009}. 

\section{Samples}
\label{sec:data}

We compiled a broad literature sample of metal column densities in the neutral ISM for 528 lines of sight and in diverse galactic environments: the MW, the LMC, the SMC, and QSO- and GRB-DLAs. We use the data collected in \citet{Konstantopoulou2022}, which includes column densities from \citet{Jenkins2009, DeCia2021, Welty2010, Phillips1982} for the MW, from \citet{Roman-Duval2021} for the LMC , from \citet{Welty2010, Tchernyshyov2015, Jenkins2017} for the SMC, from \citet{Berg2015, DeCia2018} for QSO-DLAs and from \citet{Bolmer2019} for GRB-DLAs. We additionally include column densities for the MW from 
\citet{Ritchey2023}. We cross-matched samples within the same galactic environment and combined column densities of different metals corresponding to the same line of sight. For the MW, column densities from \citet{Ritchey2023} are preferred, because they include also higher resolution data \citep[R$\sim$114,000,][]{Ritchey2023pcl}. Our final sample consists of column densities for 17 metals in their dominant ionization state, namely: \ion{O}{i}, \ion{Mg}{ii}, \ion{Si}{ii}, \ion{S}{ii}, \ion{P}{ii}, \ion{Ti}{ii}, \ion{Cr}{ii}, \ion{Fe}{ii}, \ion{Co}{ii}, \ion{Ni}{ii}, \ion{Zn}{ii}, \ion{Al}{ii}, \ion{Ge}{ii}, \ion{Kr}{i}, \ion{Cl}{ii}, \ion{Cu}{ii}, and \ion{C}{ii}.
The column densities were measured through absorption-line spectroscopy in spectra observed mainly with VLT/UVES, VLT/X-SHOOTER, HST/STIS, and HST/COS. They are homogenized to the most recent published oscillator strengths from \citet{Konstantopoulou2022} and \citet{Ritchey2023}. 

\section{Methods and results}
\label{sec:meth_res}

In the following sections, we describe our methodology. Specifically, we describe the selection of dust tracers in Sect. \ref{sec: design}, the correlations among different dust tracers in Sect. \ref{sec:correlations}, the derivation of the overall strength of depletion in Sect. \ref{sec: delta}, the usage of indirect reference dust tracers in Sect. \ref{sec: indirect}, and the determination of the dust depletion of individual elements in Sect. \ref{sec: dust_depl_elements}. We additionally provide examples ofsome  possible applications of our method in Sect. \ref{sec:stage2:application}.

\subsection{Stage 1: Designing DUNE}
\label{sec: design}

In our method, we use the correlations between all observed relative abundances of metals [X/Y] that qualify as dust tracers, to estimate the overall strength of depletion along individual lines of sight. 
A relative abundance [X/Y] qualifies as a dust tracer if it fulfills the following criteria:

\begin{enumerate}
    \item X is less refractory than Y. We assess this by calculating the difference between the refractory indices B2$_{X}$ of X and Y, which are taken from Table 4 of \citet{Konstantopoulou2022} (using the constant $\alpha_{X}$ assumption) and from \citet{Konstantopoulou2023} for P. We select relative abundances with $\Delta$B2$_{X}$ $>$ 0, to ensure that X is less refractory than Y. This criterion only gives an initial selection of [X/Y] as likely dust tracers and rejects the specular [Y/X]. No relative abundances [X/Y] were rejected based on the B2$_{X}$,  as the B2$_{X}$ coefficients result from the relative method. Because we are seeking an independent criterion instead, the dust tracers are later filtered based on the strength of their correlation (see Sect. \ref{sec:correlations}).
    
    \item X and Y follow each other in terms of the process of nucleosynthesis, to a first approximation. We categorize elements into two distinct groups depending on whether they are affected (group 1) or not (group 2) by $\alpha$-element enhancement: group 1 includes the $\alpha$-elements \citep[O, S, Si, Mg, and Ti,][]{Kobayashi2006}; whereas group 2 includes non-$\alpha$ elements (meaning all other elements that do not exhibit $\alpha$-element enhancement). Those are Fe, Cr, Ni, P, Kr, Ge, Co, Cu, Al, C, Cl, and Zn. We treated the two groups separately, namely, we did not mix group 1 with group 2 to obtain relative abundances, to ensure that our relative abundances [X/Y] are not affected by $\alpha$-element enhancements. 
    We further tested the exclusion of Zn from the analysis to ensure the robustness of the method (see Appendix \ref{excl_zn}).
\end{enumerate}

The elements are only categorized based on whether $\alpha$-element enhancement has been observed, rather than strictly by their production channels. As a result, our non-$\alpha$ element group (group 1) includes such elements as  Ge and Kr, which are produced through different channels; however, their common characteristic is that they do not show $\alpha$-element enhancement. Similarly, Ti, although it is both an Fe-peak and an $\alpha$-element, is included in group 1 with the $\alpha$-elements, because it exhibits $\alpha$-element enhancement. Here, Ti is mostly produced in core-collapse SNe \citep[e.g.,][]{Kobayashi2020}. Group 2 is a mixed group containing Fe-peak elements as well as any other non-$\alpha$ elements. Then, Ge is vastly produced in Type II SNe, and not at all in SNe Type Ia. However, it is not affected by $\alpha$-element enhancement and is included in group 2. Similarly, Kr is created 
chiefly by s-process neutron capture \citep[e.g.,][]{Sneden2008} and is not an Fe-peak element. However, it is not affected by $\alpha$-element enhancement and is included in group 2.

A reliable dust tracer is characterized by a large difference between the refractory indices of X and Y. The larger the difference in the tendencies of X and Y to deplete into dust, the larger the expected values of dust tracers [X/Y]. For example, [Si/Ti], [Zn/Fe], and [Ge/Ni] exhibit large $\Delta$B2$_{XY}$ differences, with $\Delta$B2$_{SiTi}$ = 0.92, $\Delta$B2$_{ZnFe}$ = 0.99, and $\Delta$B2$_{GeNi}$ = 0.91, respectively, and are sensitive tracers of dust. Due to the strength of these dust tracers, we used them as "reference dust tracers". On the other hand, other dust tracers are less sensitive, such as [O/Mg] with $\Delta$B2$_{OMg}$ = 0.40, while the most sensitive dust tracer in our sample is [O/Ti] with $\Delta$B2$_{OTi}$ = 1.44, but with scarce data available for O. Due to our selection criteria (defined above), reliable dust tracers are not affected by metallicity variation or nucleosynthesis effects, and any effect on them is due to dust depletion. When elements are grouped by nucleosynthetic history, meaning that they are not affected by $\alpha$-element enhancement, each set of dust tracers [X1/Y1] and [X2/Y2] is proportional to the overall strength of depletion and is, thus, correlate with the other. We identified 70 dust tracers, after applying the above selection criteria. The identified dust tracers, along with their $\Delta$B2$_{XY}$ differences, are reported in Table C.1 (available on Zenodo).

\subsection{Correlations among dust tracers}
\label{sec:correlations}

\begin{figure}
    \centering
    \includegraphics[width=\columnwidth]{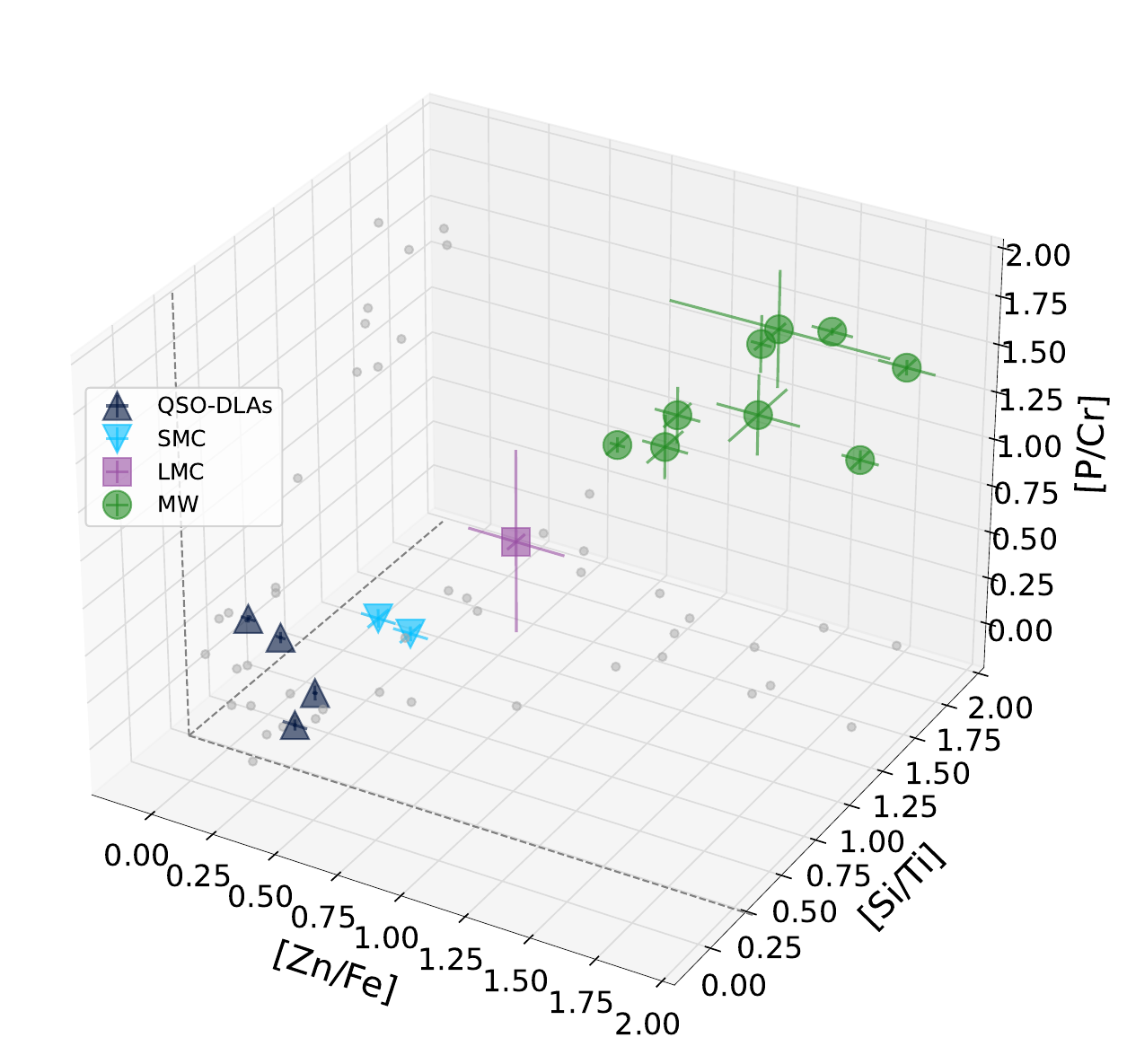}
    \caption{3D correlation between the dust tracers [Si/Ti], [Zn/Fe] and [P/Cr]. The black triangles are for QSO-DLAs,  purple squares are for the LMC,  blue triangles are for the SMC, and green circles are for the Milky Way. The gray dots are the projections of the points in each plane.}
    \label{fig:3D}
\end{figure}

We observed linear correlations between multiple dust tracers, 70 in total, considering a diversity of galactic environments (the MW, LMC, SMC, and QSO- and GRB-DLAs). In fact, all dust tracers are linearly correlated  with each other in a multidimensional space. Figure \ref{fig:3D} shows the linear correlation between three different dust tracers ([Si/Ti], [Zn/Fe], and [P/Cr]) in a three-dimensional (3D) space.

Driven by the observation that all dust tracers exhibit correlations within a multidimensional space, we found all possible correlations among relative abundances (dust tracers) and assess their correlation strength using Pearson's correlation coefficients (r). Valid correlations must meet the following criteria: 1) the correlation coefficient between the dust tracer and the reference dust tracer (r$_{XY}$) must be r$_{XY} >$ 0.6. We chose this threshold to ensure a strong correlation between the two dust tracers. 2) We selected only pairs of relative abundances with at least four common data points and with uncertainties of the relative abundances involved in each correlation less than 0.5\,dex.

\begin{figure*}
    \centering
    \includegraphics[width=\textwidth]{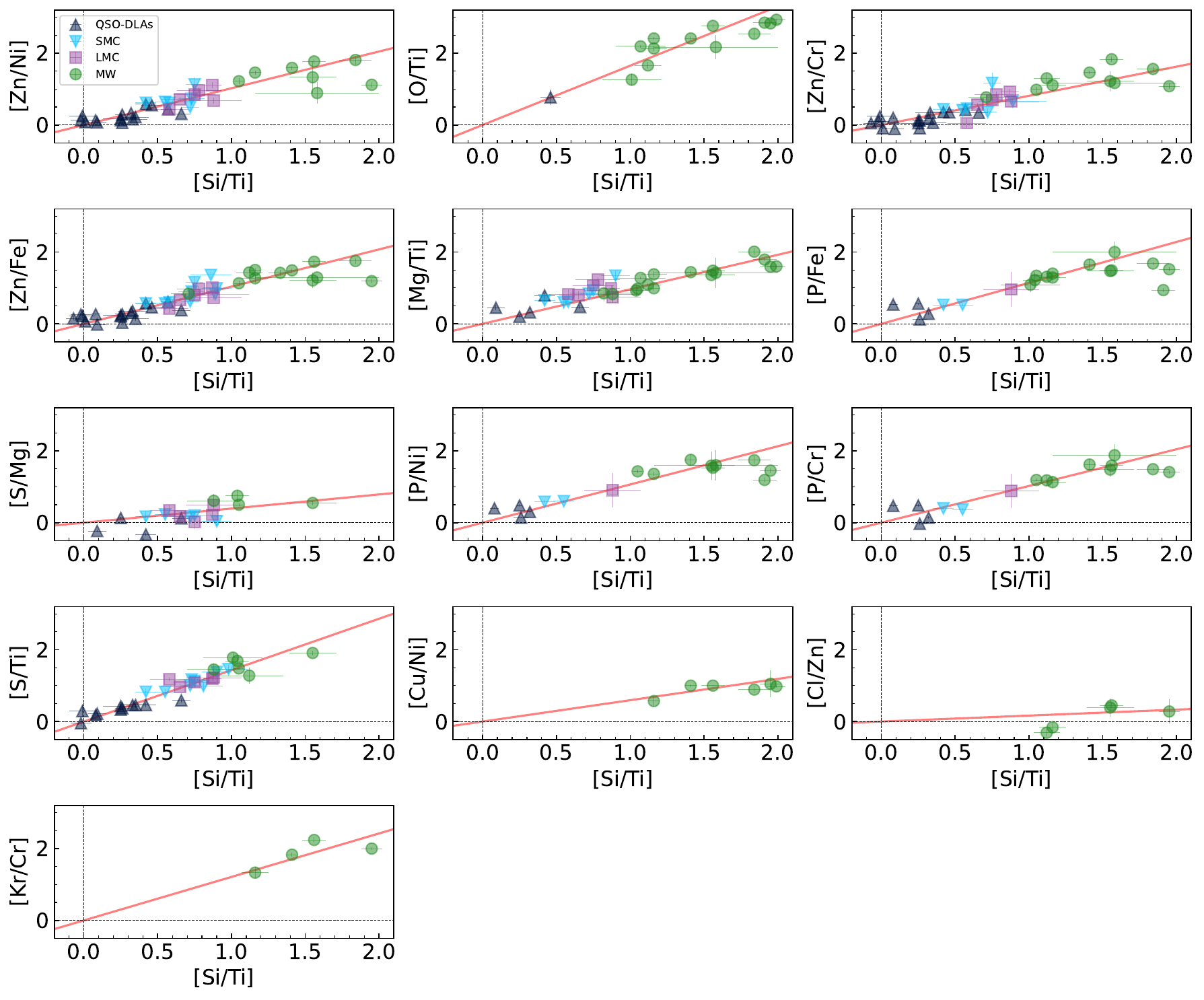}
    \caption{Correlations of different dust tracers with the primary reference dust tracer [Si/Ti]. The black triangles are for QSO-DLAs,  blue triangles for the SMC,  purple squares for the LMC and green circles for the MW. The red solid line is the linear fit to all the environments.}
    \label{fig:siti_corr}
\end{figure*}

We fit the valid correlations using orthogonal distance regression, which takes errors in both the x-axis and y-axis into account. We implement this by using the Python package scipy.odr, which determines the parameters that minimize the sum of the squared error for each data point during the regression. We fit assuming a fixed intercept at zero. This is justified by the expectation that all dust tracers should be at zero if no dust is present. Indeed, the observations agree very well with this when sufficient data is available. We chose three strong dust tracers as references to project the correlations on these most reliable axes. We assigned a priority to [Si/Ti] as our primary reference dust tracer. We expect Si and Ti to trace each other closely in nucleosynthesis, because not only they are both $\alpha$-elements, but they are also produced in the explosive environments of core-collapse SNe \citep[e.g.,][]{Nomoto2006}, as opposed to quiescent burning in their progenitors. We then assigned [Zn/Fe] as the second priority and the [Ge/Ni] as third priority reference dust tracers. We used [Ge/Ni] as a reference dust tracer instead of [P/Cr] (shown in Fig. \ref{fig:3D} due to better visualization in all the environments). While both elements are abundantly available, Ge has more consistent and reliable measurements \citep{Ritchey2023} than P does. We refer to [Zn/Fe] and [Ge/Ni] as "indirect reference dust tracers" and discuss their usage in Sect. \ref{sec: indirect}. Correlations of different valid dust tracers with the primary dust tracer [Si/Ti] and their fits are shown in Fig. \ref{fig:siti_corr}.
The parameters resulting from the fit among valid correlations between different dust tracers ([X/Y]) with respect to [Si/Ti], [Zn/Fe], and [Ge/Ni] are reported in Table 2 (available on Zenodo). Large slopes C$_{XY}$ resulting from the fit between two dust tracers denote strong dust tracers. 

\subsection{Overall strength of depletion: $\Delta$}
\label{sec: delta}

The overall strength of depletion per line of sight represents the collective contribution of dust tracers to dust depletion along a line of sight. In the hyperspace of the correlated dust tracers, the overall strength of depletion, $\Delta$, is a measure of the distance along the correlation from the origin (which is the point of no depletion). We calculate $\Delta$ with respect to [Si/Ti], and therefore it inherits its dynamical range. The overall strength of dust depletion, $\Delta$, has values that are overall similar to [Si/Ti], but are actually calculated from all the available dust tracers.

To derive $\Delta$, we calculate the sum of all the available dust tracers in each line of sight and normalize by the corresponding slopes of all the involved dust tracers with a "reference dust tracer". Each valid dust tracer [X/Y] is weighted by its correlation with a reference dust tracer. We use [Si/Ti] as our primary reference dust tracer, given its $\Delta$B2$_{X}$ strength ($\Delta$B2$_{X_{SiTi}}$ = 0.92). The overall strength of depletion for each line of sight can be estimated as

\begin{equation}
    \Delta =\frac{\sum_{i} ~[X/Y]_{i} \times W_{i} \times \frac{1}{C_{XY_{i}}}}{\sum_{i} ~ W_{i}}
    \label{eq:Delta}
,\end{equation}
where each dust tracer [X/Y]$_{i}$ is weighted by W$_{i}$ as

\begin{equation}
    W_{i} =\frac{r_{XY_{i}}}{\sigma^2} 
    \label{eq:weight}
,\end{equation}
where r$_{XY_{i}}$ is the correlation coefficient between each dust tracer with the primary reference dust tracer [Si/Ti] and C$_{XY_{i}}$ is the slope of each dust tracer with respect to the reference dust tracer. Then, $\sigma$ is given by the combined uncertainties of [X/Y] and C$_{XY_{i}}$ and is estimated as 

\begin{equation}
    \sigma = \sqrt{\left (\frac{\sigma ([X/Y])}{C_{XY}} \right)^2 + \left (\frac{[X/Y] \times \sigma(C_{XY})}{C_{XY}^2} \right)^2}
    \label{eq:sigma}
.\end{equation}
By taking into account the weights W$_{i}$ for the calculation of the overall strength of depletion (Eq. \ref{eq:Delta}), we make sure that dust tracers with large uncertainties and poor correlations have a smaller influence on the calculation. The uncertainty on the overall strength of depletion can be estimated as

\begin{equation}
    \sigma(\Delta) =(\Sigma_{i} W_{i})^{-1/2}
    \label{eq:eDelta}
.\end{equation}
The correlations of different dust tracers with the primary dust tracer [Si/Ti] are shown in Fig. \ref{fig:siti_corr}.

\subsection{Indirect reference dust tracers}
\label{sec: indirect}

In cases where [Si/Ti] was not observed for a specific line of sight and there is no correlation between the [X/Y] and [Si/Ti] (and no C$_{XY_{SiTi}}$), alternative dust tracers were used as references. Given the $\Delta$B2$_{X}$ strength of [Zn/Fe], [Ge/Ni] and their reliability as dust tracers (as described in Sect. \ref{sec: design}), we use them as indirect reference dust tracers. 
All valid dust tracers are correlated with each other and with [Si/Ti], when sufficient data are available. We calculate $\Delta$ for the cases where indirect reference dust tracers ([Zn/Fe] and [Ge/Ni]) are used, by substituting the slope C$_{XY_{i}}$ in Eq. \ref{eq:Delta} with the indirect slope C$_{XY_{\rm ind}}$. When [Zn/Fe] is used as an indirect reference dust tracer the indirect slope is estimated as

\begin{equation}
       C_{XY_{\rm ind}} = C_{XY_{ZnFe}} \times C_{ZnFe_{SiTi}}
    \label{Delta_indirect1}
,\end{equation}
where C$_{XY_{ZnFe}}$ is the slope of each dust tracer [X/Y] with the indirect reference dust tracer [Zn/Fe] and C$_{ZnFe_{SiTi}}$ is the slope of [Zn/Fe] with the primary reference tracer [Si/Ti]. Each dust tracer [X/Y] is weighted by W$_{i}$ using Eq. \ref{eq:weight}, where r$_{XY_{i}}$ is given by the indirect correlation r$_{XY_{\rm ind}}$. When [Zn/Fe] is the indirect reference dust tracer the indirect correlation is estimated as

\begin{equation}
    r_{XY_{\rm ind}} = r_{XY_{ZnFe}} \times r_{ZnFe_{SiTi}}
,\end{equation}
where r$_{XY_{ZnFe}}$ is the correlation coefficient between each dust tracer [X/Y] with the indirect reference dust tracer [Zn/Fe] and r$_{ZnFe_{SiTi}}$ is the correlation coefficient between [Zn/Fe] with the primary reference tracer [Si/Ti].

In the case where [Ge/Ni] is used as an indirect reference dust tracer (third priority dust tracer), we  use the indirect slope C$_{XY_{\rm ind}}$ instead, which is estimated as

\begin{equation}
       C_{XY_{\rm ind}} = C_{XY_{GeNi}} \times C_{GeNi_{ZnFe}} \times C_{ZnFe_{SiTi}}
    \label{Delta_indirect2}
,\end{equation}
where C$_{XY_{GeNi}}$ is the slope of each dust tracer [X/Y] with the third priority indirect tracer [Ge/Ni], C$_{GeNi_{ZnFe}}$ is the slope of [Ge/Ni] with the indirect reference tracer [Zn/Fe], and C$_{ZnFe_{SiTi}}$ is the slope of [Zn/Fe] with the primary reference tracer [Si/Ti].
The indirect correlation is estimated, in this case, as

\begin{equation}
    r_{XY_{\rm ind}} = r_{XY_{GeNi}} \times r_{GeNi_{ZnFe}}  \times r_{ZnFe_{SiTi}}
,\end{equation}
where r$_{XY_{GeNi}}$ is the correlation coefficient between each dust tracer [X/Y] with the indirect reference dust tracer [Ge/Ni], r$_{GeNi_{ZnFe}}$ is the correlation coefficient of [Ge/Ni] with the indirect reference tracer [Zn/Fe] and r$_{ZnFe_{SiTi}}$ is the correlation coefficient of [Zn/Fe] with the primary reference tracer [Si/Ti]. We adopt this approach for [Ge/Ni] as an indirect tracer due to its poor correlation with [Si/Ti], which is reflected by their very low correlation coefficient (r = 0.0021). However, [Ge/Ni]  exhibits a strong correlation with [Zn/Fe] and, subsequently, with [Si/Ti], enabling an indirect correlation of [Ge/Ni] with [Si/Ti] through [Zn/Fe]. We note that the weak correlation between [Ge/Ni] and [Si/Ti] is due to the lack of data at low levels of depletion. The correlations among different dust tracers with the primary and indirect reference dust tracers color coded by the overall strength of depletion are shown in Figs. B.2, B.3, and B.4, (available on Zenodo).

\subsection{Dust depletion of individual elements}
\label{sec: dust_depl_elements}
At this stage, we have calculated the overall strength of depletion per line of sight, which represents the collective contribution of the depletion of all available elements.  We can then focus on determining the dust depletion of individual elements, utilizing the derived overall strength of depletion and the observed relative abundances of metals. Given that the overall strength of depletion ($\Delta$) considers the contribution of all the elements, there must be a correlation between $\Delta$ and the depletion of individual elements.

The dust depletion of element X, $\delta_{X}$, can be derived from a relative abundance of an element that is highly volatile, and correcting for the depletion of the volatile element. In our sample, Kr is the least depleted metal \citep[e.g.,][]{Konstantopoulou2022}. Because we measure Kr column densities from \ion{Kr}{i}, and we do not actually expect Kr to be depleted, we assume that Kr has no depletion. We discuss this in more detail in Sect. \ref{sec: krypton}. Then we can derive the dust depletion of each element as

\begin{equation}
    \delta_{X} - \delta_{Kr} = [\rm {X/Kr}] = B_{X} \times \Delta \sim \delta_{X}
    \label{eq: xkr_delta}
,\end{equation}
where B$_{\rm{X}}$ is the slope of the linear fit unique to each element X and $\Delta$ is the overall strength of depletion derived from Eq. \ref{eq:Delta}. The coefficients B$_{\rm{X}}$ are reported in Table \ref{xkr_coefficients}.

\begin{figure*}
    \centering
    \includegraphics[width=\textwidth]{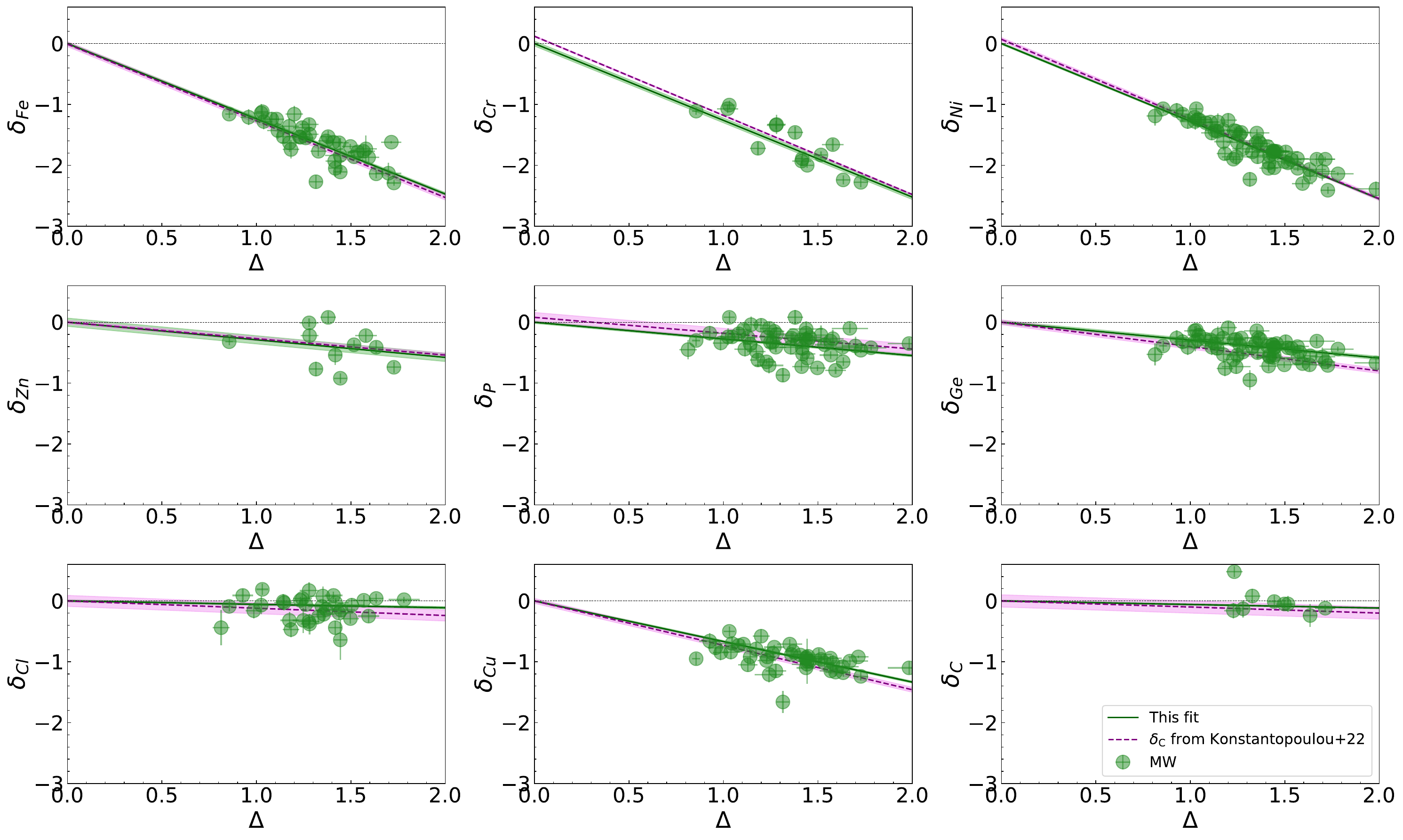}
    \caption{Dust depletion of element X against the overall strength of depletion $\Delta$ for lines of sight in the MW. The green solid line shows the linear fit to the data and the dashed purple line shows the depletion of element X ($\delta_{\rm{X}}$) from \citet{Konstantopoulou2022}. The colored surfaces show the uncertainty areas of the fits.}
    \label{fig:xkr_delta}
\end{figure*}

Figure \ref{fig:xkr_delta} shows the dust depletion of element X ($\delta_{X}$) (for X: Cl, P, Ni, Cu, Cr, Zn, Ge, Fe, and C) with respect to the overall strength of depletion $\Delta$ for the MW. We observe a linear relation between $\delta_{X}$ and $\Delta$ among all the elements that were studied and fit the data with orthogonal distance regression.

\section{Stage 2: Application of the method}
\label{sec:stage2:application}

In this section, we provide practical guidance on how to apply the methodology described above to derive the overall strength of depletion, $\Delta$, and analyze the abundance patterns, metal patterns, and relative metal patterns.

\subsection{Abundance patterns when $\rm{H}$ is known}
\label{sec: stage2:abundance_patterns}

An analysis of the abundance patterns can be carried out when H is known. It can be used to derive the overall amount of dust depletion, metallicity of the gas, and eventual deviations from the metal pattern of specific elements, for example, due to nucleosynthesis effects such as $\alpha$-element enhancements from core-collapse SNe.

\begin{figure}
    \centering
    \includegraphics[width=0.5\textwidth]{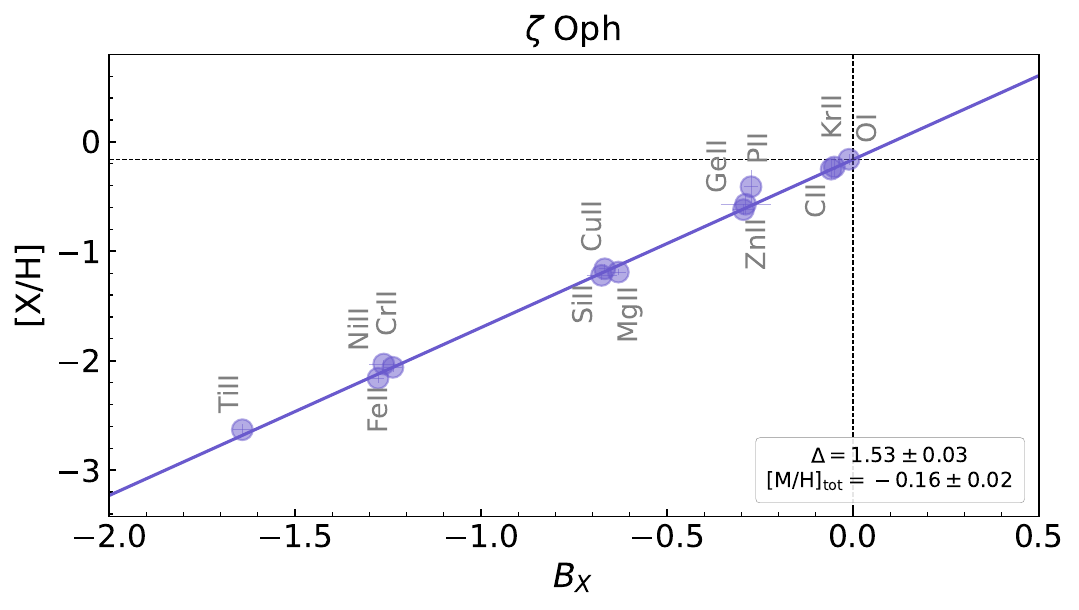}
    \caption{Abundances [X/H] with respect to the coefficients B$_{X}$ (abundance pattern) resulting from Eq. \ref{eq: xkr_delta} for $\zeta$\,Oph in the MW. The solid line is the fit to the points, the slope of which gives the overall strength of depletion $\Delta$ along the line of sight. The y intercept gives the total metallicity [M/H]$_{\rm{tot}}$.}
    \label{fig:XH_Cx_example}
\end{figure}

This analysis is analogous to figure 3 of \citet{DeCia2024}, but using the DUNE coefficients. Figure \ref{fig:XH_Cx_example} shows an example of the study of an abundance pattern. The abundance pattern shows the observed abundances [X/H] (on the y-axis) with respect to the tendencies of each metal to deplete into dust, which is quantified by the parameter B$_{X}$ (x-axis), which we call the DUNE refractory index. The total metallicity [M/H]$_{\rm{tot}}$ and the overall strength of depletion $\Delta$ can be derived as

\begin{equation}
    [X/\rm{H}] = [M/H]_{\rm{tot}} + \Delta \times B_{X}
    \label{eq: abundance_patterns}
,\end{equation}
which can be written as a linear relation, $y = \alpha + b\,x$, where the values of B$_{X}$ and [X/H] are the known \textit{x} and \textit{y} data (tabulated and observed, respectively). The normalization of the linear fit gives the metallicity of the gas ($\alpha = [M$/H]$_{\rm{tot}}$), the slope gives the overall strength of depletion (b = $\Delta$), and any deviations on individual elements can be identified, are due to nucleosynthesis effects. The coefficients B$_{\rm{X}}$, are reported in Table \ref{xkr_coefficients}. In principle, one can calculate $\Delta$ already from one relative abundance [X/Y] of a dust tracer, as described in Sect. \ref{sec: stage2:relative_metal_patterns}.

\subsection{Metal patterns  when $\rm{H}$ is unknown}
\label{sec: stage2:metal_patterns}

\begin{figure}
    \centering
    \includegraphics[width=0.5\textwidth]{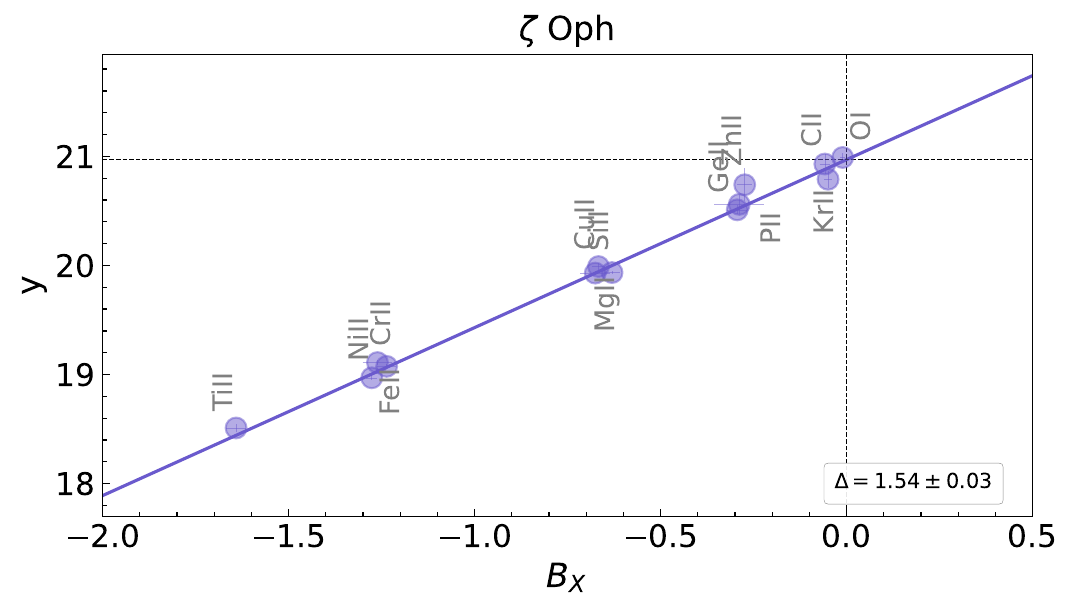}
    \caption{Metal pattern for $\zeta$\,Oph in the MW. The solid line is the fit to the points, the slope of which gives the overall strength of depletion $\Delta$ along the line of sight.}
    \label{fig: metal_pattern}
\end{figure}

The analysis of the metal patterns can be done in case H is unknown, including for individual components of the line profiles \citep{Ramburuth-Hurt2023}. The metal patterns can be useful to derive the overall amount of dust depletion, $\Delta$, and to identify eventual deviations from the metal pattern of specific elements, for example, due to nucleosynthesis effects, such as $\alpha$-element enhancement from core-collapse SNe. The metal patterns can be described by rewriting Eq. \ref{eq: abundance_patterns}, as follows

\begin{equation}
    y = a + B_{X} \times \Delta
    \label{eq: metal_pattern0}
,\end{equation}
where 
\begin{align}
    y &= \log\,N(X) - X_{\odot} + 12, \\
    a &= [M/\rm{H}] + \log\,N(\rm{H}).
    \label{eq: metal_pattern1}
\end{align}
In this case, \textit{y} shows the equivalent metal column densities of metals (abundances normalized by their equivalent solar abundance), rather than the metal abundances, as shown in Sect. \ref{sec: stage2:abundance_patterns}. Figure \ref{fig: metal_pattern} shows the metal pattern for $\zeta$\,Oph in the MW. The y-intercept of the linear fit to the data gives the equivalent metal column density, a = [M/$\rm{H}] + \log\,N(\rm{H}$) -- or the total amount of metals after correcting for dust depletion -- and the slope gives the overall strength of depletion, $\Delta$. Deviations of individual elements from the linear fit, are due to nucleosynthesis effects, such as $\alpha$-element enhancement. These deviations (i.e., the distance in the y-axis of the observed data  from the main linear fit) can be used to quantify the alpha-element enhancements in the ISM or potential other nucleosytnhesis signatures.

\subsection{Relative metal patterns}
\label{sec: stage2:relative_metal_patterns}
 
\begin{figure}
    \centering
    \includegraphics[width=0.5\textwidth]{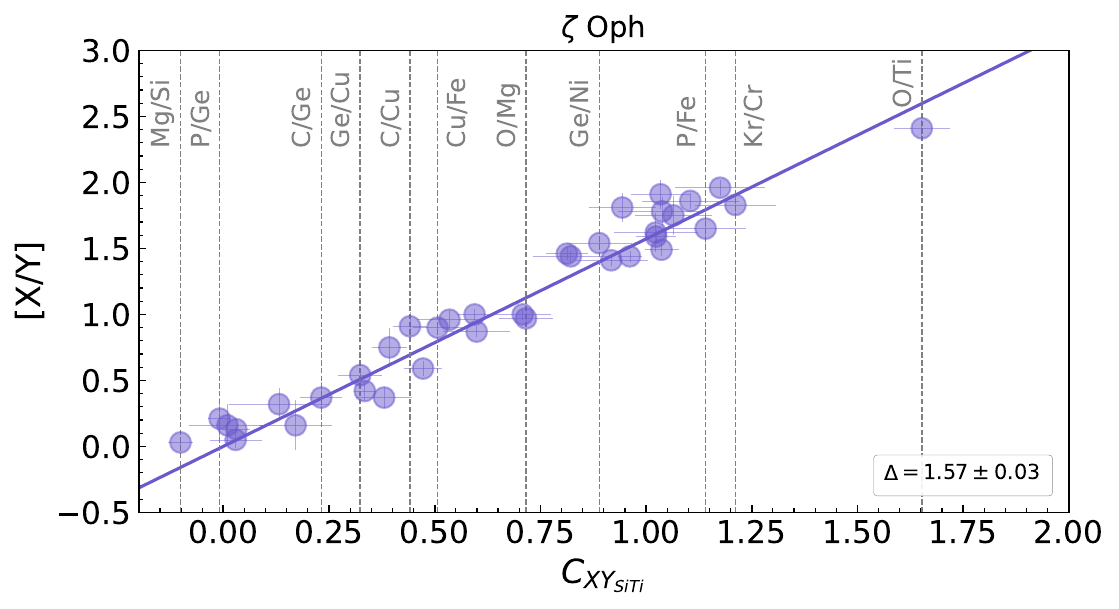}
    \caption{Relative abundances [X/Y] with respect to their corresponding slopes with the dust tracer [Si/Ti] (C$_{\rm XY_{\rm SiTi}}$) for $\zeta$\,Oph in the MW. The solid line is the fit to the points, the slope of which gives the overall strength of depletion $\Delta$ along the line of sight. The vertical dashed lines show the dust tracers that can be found in the C$_{\rm XY_{\rm SiTi}}$ range.}
    \label{fig:XY_Bxy_example}
\end{figure}

The analysis of the relative metal patterns can only be used to determine dust depletion. This means that the relative metal patterns should not be used in cases where potential deviations due to nucleosynthesis effects might occur. Instead, it is recommended to use [X/Y] ratios that are pure dust tracers, avoiding mixtures of elements with different nucleosynthetic origins (e.g., $\alpha$ and non-$\alpha$ elements).
The overall strength of depletion $\Delta$ can then be calculated as

\begin{equation}
    \Delta  = \frac{[X/Y]}{C_{XY}} 
    \label{stage2_Delta}
,\end{equation}
using the observed relative abundances [X/Y] and the coefficients C$_{XY}$, which are reported in Table 2. Equation \ref{stage2_Delta} is technically Eq. \ref{eq:Delta} if applied to one dust tracer and for W$_{i}$ = 1.
In Fig. \ref{fig:XY_Bxy_example}, we provide an example of how the user can practically obtain $\Delta$. It shows the relative abundances [X/Y] against their slopes with the reference dust tracer [Si/Ti] (C$_{\rm XY_{\rm SiTi}}$) for $\zeta$\,Oph in the MW. The slope of the linear fit to the data gives the overall strength of depletion. $\Delta$.

\section{Discussion}
\label{sec: discussion}

\subsection{Strengths and limitations}
\label{sec: limitations}

While our method is one of the most robust methodologies available so far, since it combines strengths from previous methods and overcomes their weaknesses, it also faces some limitations. One of the key strengths of our approach is its ability to derive dust depletion without relying on assumptions regarding $\alpha$-element enhancement, metallicity, or dust behavior, thus offering a more reliable determination of both the overall strength of depletion and the dust depletion of individual metals. Additionally, our method can be applied to individual gas clouds, allowing for a more detailed analysis of dust depletion within specific components of the ISM.

However, one limitation that our method is facing, lies in the use of Kr for the determination of the depletion of individual elements through Eq. \ref{eq: xkr_delta}. To obtain the depletion of X through the relative abundance [X/Kr], without applying any additional corrections, the depletion of Kr has to be consistent with zero. We estimated the depletion of Kr, from the slope of the linear relation [Kr/H] = B$_{Kr} \times \Delta$, to be B$_{Kr}$ = -0.05 $\pm$ 0.01. However, there is a large scatter in the data, as seen in Fig. \ref{fig:xh_tracers}, which might be an effect of metallicity variations in the gas. In addition, the data span a short dynamical range in the x-axis ($\Delta$), not allowing for an accurate determination of the slope.

Another limitation is the availability of metals. Because we are using relative abundances of metals, at least four different metal measurements need to be available in each line of sight, in order to make it possible to obtain correlations among them and consequently the coefficients for their fits. Therefore, we have not been able to provide coefficients for some metals, due to a lack of sufficient measurements, for example, S, Cl, P, Al, Cu, and Co.

We tested whether our methodology is affected by lines of sight that have broken linear relations in their abundance patterns due to multiple gas components with different depletions, as found in \citet{DeCia2021} and \citet{Ritchey2023}. To test this we exclude from the analysis lines of sight that exhibit "upturns" of the volatile elements in the \citet{Ritchey2023} sample. The derived depletions are not significantly affected by the presence of these lines of sight, thus, we include them here.

Therefore, while our method provides the most solid way to derive dust depletions so far, some limitations are still present mainly due to the availability and quality of metal column density measurements.

\subsection{Comparison with previous methods}
\label{sec: comparisons}

Here, we give some examples of how our results compare with the results obtained with the relative method \citep{DeCia2021} and with the F$_{\ast}$ method \citep{Jenkins2009, Ritchey2023}. The $\Delta$ determined in DUNE is related with F$_{\ast}$ by the linear relation F$_{\ast}$ = 0.86 $\times$ $\Delta$ - 0.54, based on F$_{\ast}$ data from \citet{Ritchey2023}. The relation between $\Delta$ and [Zn/Fe] is [Zn/Fe] = 1.05 $\times$ $\Delta$ - 0.03, based on the data of this work. We convert our $\Delta$ to [Zn/Fe] and to F$_{\ast}$ to be able to compare the results of the three methods on a similar scale.

Taking a line of sight that is in common in the three methods, for example, $\epsilon$\,Per in the MW, we can compare the values on the overall strength of depletion and the gas-phase metallicity. With DUNE we find the overall strength of depletion, found with the abundance pattern, of $\Delta$ = 1.26 $\pm$ 0.09. This corresponds to [Zn/Fe] = 1.29 $\pm$ 0.02. With the relative method \citet{DeCia2021} find [Zn/Fe]$_{\rm{fit}}$ = 1.24 $\pm$ 0.13, which is consistent with the [Zn/Fe] we calculate using its relation to $\Delta$. Converting to F$_{\ast}$ we derive F$_{\ast}$ = 0.54 $\pm$ 0.17, which is consistent with the F$_{\ast}$ = 0.68 $\pm$ 0.04 derived in \citet{Jenkins2009}. For the metallicity, we find [M/$\rm{H]_{tot}}$ = -0.14 $\pm$ 0.09, while \citet{DeCia2021} find [M/$\rm{H]_{tot}}$ = -0.18 $\pm$ 0.13. Therefore, our results on the depletion and metallicity for $\epsilon$\,Per are consistent with the results from \citet{Jenkins2009, DeCia2021}, within the uncertainties. 

For $\zeta$\,Oph, which is featured in Figs. \ref{fig:XH_Cx_example}, \ref{fig: metal_pattern}, \ref{fig:XY_Bxy_example}, in DUNE we find using the abundance pattern an overall strength of depletion $\Delta$ = 1.53 $\pm$ 0.03. This corresponds to F$_{\ast}$ = 0.78 $\pm$ 0.15. \citet{Jenkins2009} finds F$_{\ast}$ = 1.05 $\pm$ 0.02 while \citet{Ritchey2023} find F$_{\ast}$ = 0.85 $\pm$ 0.05. Our estimate of the overall strength of depletion is in better agreement with \citet{Ritchey2023}. In DUNE we find a metallicity [M/$\rm{H]_{tot}}$ = -0.16 $\pm$ 0.03, assuming a fit to all the metals. In \citet{Ritchey2023} they find using the F$_{\ast}$ method [M/$\rm{H]_{tot}}$ = -0.07 $\pm$ 0.06, fitting all the metals, consistent within the uncertainties with our findings. This line of sight was not studied in \citet{DeCia2021}.

\subsection{Coefficients for mixed groups}
\label{sec: a_elements_coef}

As previously highlighted, our methodology strictly avoids mixing groups of elements when defining dust tracers. This means that we never include relative abundances such as [group1/group2], ensuring that $\alpha$-elements are not mixed with non-$\alpha$ elements. This is to avoid the influence of $\alpha$-element enhancement, in the relative abundances (dust tracers). Alternatively, corrections for nucleosynthesis effects would be necessary, as described in \citet{DeCia2016} and \citet{Konstantopoulou2022}. This would require the assumption of nucleosynthetic curves for each environment. However, the exact shape of the nucleosynthetic curves, and especially the position of the $\alpha$-element knee, is uncertain or poorly constrained for QSO-DLAs and for the Magellanic Clouds. A first determination of the distribution of the $\alpha$-element enhancement in QSO-DLAs is presented in \citet{Velichko2024}.

For the ease of use of the reader, particularly in cases where only a limited number of elements are available, we provide coefficients resulting from the fit $\delta_{X}$ = [X/Kr] =
B$_{X}$ $\times$ $\Delta$, shown in Fig. B.5 (available on Zenodo), where X is an $\alpha$-element.
This is done only for the MW data, where nucleosynthesis effects are not present. Specifically, coefficients are provided for correlations of O, Mg, Si and Ti with non-$\alpha$ elements. This extended set of coefficients ensures that our method can be applied even when only few elements are observed. The coefficients for correlations between mixed elements are reported in Table C.2 (available on Zenodo).

\subsection{Krypton}
\label{sec: krypton}

Krypton being a noble gas, remains chemically inert and does not easily create bonds with other elements. As a result, Kr is primarily found in the gas phase of the ISM and it is hard to see how it could be incorporated into dust grains. However, it is important to note that the ISM is a complex environment with a variety of physical processes occurring, including the formation and destruction of dust grains. While Kr may not directly form dust itself, it could potentially be indirectly associated with dust through other processes, which may not yet be understood. 

In our method, we aim to avoid assumptions that can compromise the validity of our results. Therefore, even though it seems unlikely that Kr depletes into dust, due to its nature, we examined the possibility of slight depletion. In our methodology, we use relative abundances rather than metal abundances to estimate dust depletion of different elements. This approach helps to avoid complications that can arise from metallicity variations or nucleosynthesis effects, such as $\alpha$-element enhancement. However, Kr being highly volatile, noble gas and created by s-process neutron capture \citep{Sneden2008}, it remains insensitive to variations of low-metallicity gas and is not affected by nucleosynthesis effects. This is demonstrated in the abundance patterns observed by \citet{DeCia2021, Ritchey2023}, where deviations from linearity, or upturns, of volatile elements in the MW indicate the complex nature of the ISM, comprising a mix of gases with different metallicities and dust depletions. In \citet{DeCia2024}, they show that these upturns of the volatile elements arise in the presence of high-metallicity gas, indicating that volatiles are less sensitive to variations in low-metallicity gas. Additionally, our analysis of Kr depletion focuses on the MW, where nucleosynthesis effects are not present. While variations of low-metallicity gas do not affect the abundance of Kr, high-metallicity gas can influence its abundance.

\begin{figure*}
    \centering
    \includegraphics[width=\textwidth]{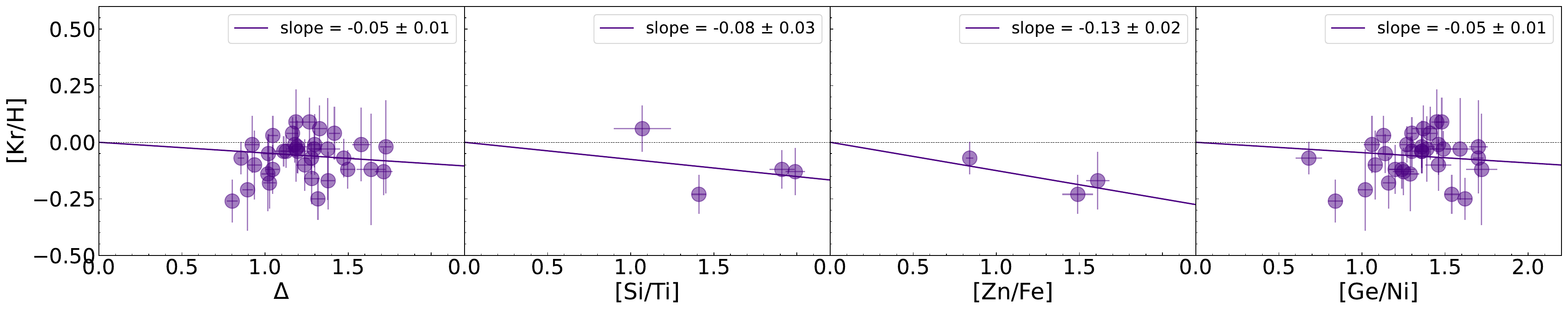}
    \includegraphics[width=\textwidth]{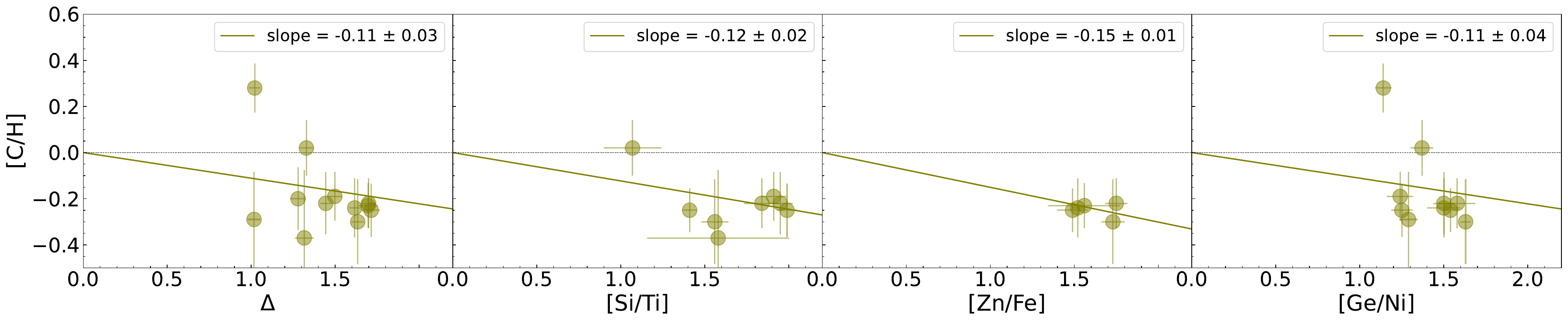}
    \includegraphics[width=\textwidth]{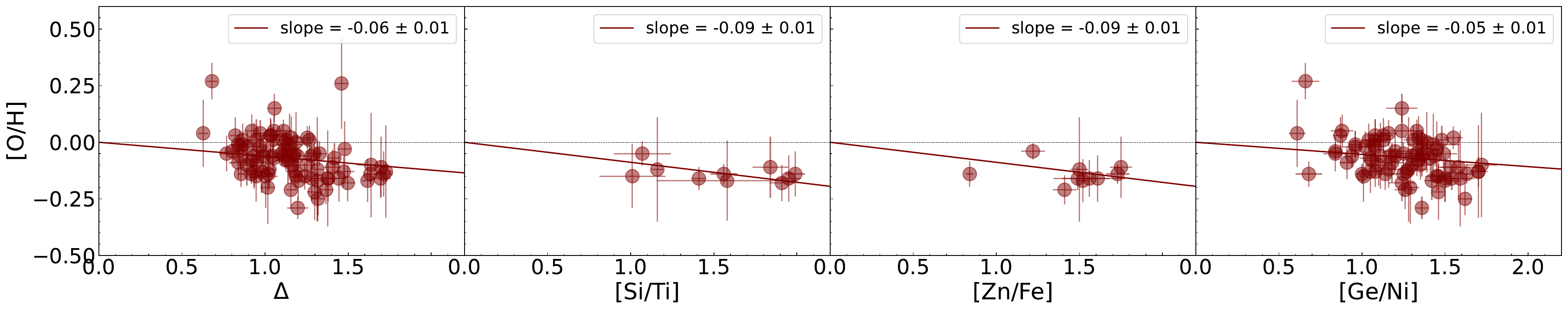}
    \caption{Abundances with respect to the overall strength of depletion $\Delta$ and the dust tracers [Si/Ti], [Zn/Fe], and [Ge/Ni] for MW data. Top panel: [Kr/H]. Middle panel: [C/H]. Bottom panel: [O/H].
    The solid lines shows the linear fit to the MW data. The slope of the fits gives the depletions of Kr, C and O respectively. ($\delta_{\rm{Kr}}$, $\delta_{\rm{C}}$, and $\delta_{\rm{O}}$).}
    \label{fig:xh_tracers}
\end{figure*}

We investigate whether there is independent evidence for the depletion of Kr by studying the relation between the abundance of Kr ([Kr/H]) with respect to the overall strength of depletion, $\Delta$, and the dust tracers [Si/Ti], [Zn/Fe], and [Ge/Ni] in the MW (as shown in Fig. \ref{fig:xh_tracers}). The top panel of Fig. \ref{fig:xh_tracers} shows the abundance of Kr ([Kr/H]) with respect to $\Delta$ and the dust tracers [Si/Ti], [Zn/Fe], and [Ge/Ni]. We fit the data using orthogonal distance regression, with the resulting slopes labeled in Fig. \ref{fig:xh_tracers}. The average slope among the different panels of Fig. \ref{fig:xh_tracers} is -0.09 $\pm$ 0.04. This is consistent, within the uncertainties, with the slope obtained from the fit with $\Delta$ (see Sect. \ref{sec: dust_depl_elements}). The slope obtained from [X/Kr] = B$_{\rm{Kr}}$ $\times$ $\Delta$ is also consistent, within the uncertainties, with the result from \citet{Konstantopoulou2022}, who reported B2$_{\rm{Kr}}$ = -0.04 $\pm$ 0.09. \citet{Jenkins2009} finds, using the F$_\ast$ method, that the depletion of Kr get progressively stronger as F$_\ast$ increases (A$_{\rm{Kr}}$ = -0.17 $\pm$ 0.10) and \citet{Ritchey2018} find the same slope but with an improved significance level in their measurement (A$_{\rm{Kr}}$ = -0.17 $\pm$ 0.06). A$_{\rm{Kr}}$ corresponds to B2$_{\rm{Kr}}$ = -0.18 $\pm$ 0.06 using the relation B2$_{X}$ = 1.05 $\times$ A$_{X}$. \citet{Jenkins2019} attempts to explain this puzzling depletion of Kr, although only marginally significant, with various scenarios. One explanation given by \citet{Jenkins2019} is that Kr atoms could attach more strongly to surfaces if those surfaces were hit by energetic protons, which create free radicals and broken bonds that can chemically attract noble gas atoms, such as Kr \citep{Hohenberg2002}. Ionizing radiation might also create such strong binding sites. This would mean that the Kr depletion is influenced by the presence of dust grains but also by higher levels of UV radiation, which can make surfaces more chemically active. As discussed in \citet{Jenkins2009} laboratory experiments and studies of meteoritic materials suggest that noble gases can bind to certain compounds or charged molecular complexes under specific conditions, but these mechanisms do not fully explain the observed depletions in the ISM. Additionally, noble gas atoms can bind to positive ions and charged molecular complexes, such as $\rm{H^{+}}$, $\rm{H^{+}_{2}}$, and $\rm{H^{+}_{3}}$. This phenomenon has been observed for interstellar $\rm{ArH^{+}}$ in spectra, however, the concentrations have typically been low. It is unlikely that such 
molecular complexes cause significant depletions of Kr.
Another possibility is that \ion{Kr}{i}, as a neutral species, can easily be ionized. In that case, Kr would be ionized more efficiently than hydrogen, lowering the apparent gas-phase 
[Kr/H] in the warm phase of the ISM that we are probing, where most singly ionized species lie. However, this would result in a roughly flat trend of the Kr abundance [Kr/H] with log\,N(\ion{H}{I}). Instead, we observed a slightly decreasing trend of [Kr/H] with increasing log\,N(\ion{H}{I}), which suggests little or no ionization. This is because, as the total hydrogen column density decreases in the neutral gas, Kr abundance appears to increase if Kr is not depleted into dust.
In addition, we do not observe a trend between [O/Kr] with respect to log\,N(\ion{H}{i}). Figure B.6 (available on Zenodo) shows the relations between [Kr/H] and [O/Kr] with log\,N(\ion{H}{I}). A suitable alternative reference instead of \ion{Kr}{I} could be \ion{Ar}{I}, which is also a noble gas and not depleted into dust grains. However, \citet{Jenkins2013} demonstrated that \ion{Ar}{I} can still be ionized in the partly ionized ISM. In addition, \ion{Ar}{I} is not sufficiently measured in the Milky Way, while \ion{Kr}{I} has been measured in several lines of sight in our sample.

A mild depletion of Kr would have a small systematic effect on our estimate of the depletion of all metals, which is relatively more important for volatile elements. We can simply subtract it from the B$_{\rm{X}}$ coefficients as (B$_{\rm{X}}$ - B$_{\rm{Kr}}$, where B$_{\rm{Kr}}$ = -0.09 $\pm$ 0.04). However, we do not consider this apparent Kr depletion significant, since it is not supported by other observed trends. On the contrary, our results on the depletion of individual elements, without the need to assume Kr depletion, agree very well with previous studies \citet{DeCia2016, Jenkins2009, Konstantopoulou2022}. The behavior of Kr remains puzzling and further observational evidence would be needed to resolve the discrepancies.

\begin{table}[!ht]
\caption{Coefficients B$_{\rm{X}}$ resulting from the linear fit $\delta_{\rm{X}}$ = B$_{\rm{X}}$ $\times$ $\Delta$ shown in Figs. \ref{fig:xh_tracers}, \ref{fig:xkr_delta}, and B.5 and the degrees of freedom $\nu$.}
\label{xkr_coefficients}
\centering
\begin{tabular}{lcccc}
\hline
 X &  B$_{\rm{X}}$ &  $\nu$ \\ \hline \hline
 Kr & -0.05 $\pm$ 0.01 &   33\\
Cl & -0.06 $\pm$ 0.02 &   30 \\
 C & -0.07 $\pm$ 0.02 &    5 \\ 
 P & -0.27 $\pm$ 0.02 &   51 \\
 Zn & -0.29 $\pm$ 0.07 &    9 \\
Ge & -0.29 $\pm$ 0.03 &   61 \\
Cu & -0.67 $\pm$ 0.02 &   40 \\
Fe & -1.24 $\pm$ 0.02 &   37 \\  
Cr & -1.26 $\pm$ 0.04 &   12 \\
Ni & -1.28 $\pm$ 0.02 &   59 \\
 \hline & $\alpha$-elements$^a$ & \\ \hline  
 O & -0.06 $\pm$ 0.01 &   61 \\
 Mg & -0.63 $\pm$ 0.02 &   63 \\
Si & -0.68 $\pm$ 0.04 &    8 \\
Ti & -1.64 $\pm$ 0.03 &   39 \\ \hline
\end{tabular}
\flushleft
{\footnotesize{{\bf Note:} $^a$ We provide coefficients for a/non-$\alpha$ group elements only for the MW and only for the convenience of the user (see Sect. \ref{sec: a_elements_coef}). We do not use mixed groups for the development of the method.}}
\end{table}

\subsection{Carbon and oxygen}
\label{sec: carbon_oxygen}

Carbon plays a significant role in dust composition, with considerable amounts found in dust grains. The bulk of C is found in the form of submicron-sized carbon dust grains, such as amorphous carbon or graphite \citep{Mishra2017}, but also in the form of polycyclic aromatic hydrocarbons (PAHs). PAHs are the main candidates for the origin of the 2175$\,\AA$ extinction feature, observed in the MW \citep[e.g,][]{Fitzpatrick&Massa1986ApJ, Gordon2003, Valencic2004, Gordon2023, Lin2023}, in local galaxies \citep[e.g,][]{GordonClayton1998, Gordon1999, Gordon2003, MaizApellaniz2012}, but also in distant galaxies \citep[e.g.,][]{Motta2002, York2006, Eliasdottir2009, Zafar2011, Witstok2023}. While different studies on dust grains may differ regarding the composition and form of dust grains, they all require the presence of carbon dust, for example in the form of graphite or PAH \citep[e.g.,][]{Mattsson2019, Roman-Duval2022a, Konstantopoulou2024}. Additionally, most models require carbon to explain the 2175$\,\AA$ extinction feature \citep[e.g.,][]{LiDraine2001, Draine2003}. From the fit in Fig. \ref{fig:xkr_delta} we find a slope B$_{C} = -0.06\pm 0.02$, which is consistent within the uncertainties with \citet{Konstantopoulou2022}, who find B2$_{C} = -0.10 \pm 0.10$ and \citet{Jenkins2009} who find A$_{C}$ = -0.10 $\pm$ 0.23, which corresponds to B2$_{C}$ = -0.11 $\pm$ 0.23. 
We exclude from the fit the point at $\delta_{C}$ = 0.51 $\pm$ 0.09. This highly positive $\delta_{C}$ arises from a high C column density. For this line of sight, the measurement of the C column density is done in \citet{Sofia2004} and may be overestimated because of the noise structure in the data. Thus, we consider this measurement unreliable and exclude it from the fit.

We estimate the amount of C in dust in particles per million (ppm) in the case of maximum overall strength of depletion $\Delta$. Considering that B$_{C}$ = -0.07 $\pm$ 0.02 we estimate the maximum amount of C depletion as $\delta_{C_{max}}$ = B$_{C}$ $\times \Delta_{C_{max}}$. Then the amount of C in dust in terms of particles per million is estimated as [C/H]$_{\rm{dust}}$=[C/H]$_{\rm{ISM}}$(1-10$^{\delta_{C_{max}}}$) = 82 $\pm$ 31\,ppm, where we adopt [C/H]$_{\rm{ISM}}$ = 339 $\pm$ 39\,ppm for C from \citet{Zuo2021} and \citet{Hensley2021}. This is consistent with [C/H]$_{\rm{dust}}$ estimated at maximum amount of dust for the Milky Way in \citet{Konstantopoulou2024} and with the dust-phase C abundance estimated by \citet{Cardelli1996}.

The middle panel of Fig. \ref{fig:xh_tracers} shows the gas-phase observed C abundance ([C/H]) with respect to $\Delta$ and the dust tracers [Si/Ti], [Zn/Fe], [Ge/Ni]. We fit the data resulting in the slopes labeled in Fig. \ref{fig:xh_tracers}, which represent the depletion of C. The average slope of the distribution of [C/H] with respect of different dust tracers (see Fig. \ref{fig:xkr_delta}) is -0.13 $\pm$ 0.05 and it is consistent with the slope obtained from the fit with $\Delta$ within the uncertainties (see Sect. \ref{sec: dust_depl_elements}). Because C is very lightly depleted, a small change in the depletion of Kr can make a bigger difference in the [C/Kr] relative abundance, than for strongly depleted elements.

Oxygen plays a crucial role in ISM dust composition, especially as a component of silicates, contributing significantly to the overall dust mass. However, using Eq. \ref{eq: xkr_delta} we find that B$_{\rm{O}} = -0.01 \pm 0.01$. This is contradicting what is already known about the role of O in dust composition. Indeed, dust depletion studies, such as those presented by \citet{Jenkins2009, DeCia2016, Konstantopoulou2022}, have revealed substantial depletions of O, with values ranging from -0.09 to -0.23. \citet{Jenkins2009} found A$_{\rm{O}} = -0.23$  $\pm$ 0.05 (corresponding to B$_{\rm{O}}$ = -0.24 $\pm$ 0.05), whereas \citet{DeCia2016} found B2$_{\rm{O}} = -0.15$  $\pm$ 0.09 and \citet{Konstantopoulou2022} found B2$_{\rm{O}} = -0.20$ $\pm$ 0.05.

The bottom panel of Fig. \ref{fig:xh_tracers} shows the O abundance ([O/H]) with respect to $\Delta$ and the dust tracers [Si/Ti], [Zn/Fe], [Ge/Ni]. We fit the data resulting in the slope B$_{\rm{O}}$, which represents the depletion of O ($\delta_{\rm{O}}$).  
The average slope of the distribution of [O/H] with respect to different dust tracers (see Fig. \ref{fig:xh_tracers}) is -0.08 $\pm$ 0.03. This is consistent, within the uncertainties, with the B$_{\rm{O}}$ obtained from the fit with $\Delta$ (see Sect. \ref{sec: dust_depl_elements}). This is also consistent within the uncertainties with \citet{DeCia2016} and is lower than the results from \citet{Jenkins2009, Konstantopoulou2022}. Therefore, Fig. \ref{fig:xh_tracers} provides a better representation of the depletion of O than Fig. B.5. Because O is mildly depleted, even a small change in the depletion of Kr can make a bigger difference in the [O/Kr] relative abundance, than for strongly depleted elements. 

However, the depletion of Kr into dust grains is highly unlikely, due to its noble gas nature, and the bonding of gas-phase atoms into solids is not observed in the ISM. The constant [O/Kr] with $\Delta$ suggests that, while O is known to deplete into dust grains, this depletion is balanced out by other processes in the ISM that maintain the [O/Kr] ratio stable throughout different amounts of dust. This trend remains puzzling, and more data and analysis are needed to understand the underlying causes of the [O/Kr] trend and the behavior of O and Kr in the complex ISM. Although there is uncertainty in the nature of the [O/Kr] - $\Delta$ trend which hints to $\delta_{\rm{O}}$ = -0.01 $\pm$ 0.01, the observed oxygen abundance ([O/H]) with respect to the amount of dust (see Fig. \ref{fig:xh_tracers}), imply a larger depletion of O with an average $\delta_{\rm{O}}$ = -0.08 $\pm$ 0.03. Additional measurements of O and/or Kr would be necessary to refine the determination of the B$_{\rm{O}}$ coefficient.

We estimate the O abundance in dust in terms of particles per million, assuming maximum overall strength of depletion $\Delta_{max}$ and estimating the maximum depletion of O as $\delta_{\rm{O}_{max}}$ = B$_{\rm{O}} \times \Delta_{max}$, where B$_{\rm{O}}$ = -0.06 $\pm$ 0.01 is slope of [O/H] with $\Delta$ shown in the bottom panel of Fig. \ref{fig:xh_tracers}. We find [O/H]$_{\rm{dust}}$ = 141 $\pm$ 60\,ppm estimated as [O/H]$_{\rm{dust}}$ = [O/H]$_{\rm{ISM}} \times (1-10^{\delta_{\rm{O}}})$, where [O/H]$_{\rm{ISM}}$ = 589 $\pm$ 68\,ppm from \citet{Zuo2021} and \citet{Hensley2021}. This means that if we assume the maximum amount of depletion, 141 particles per million of O are expected to be in dust, corresponding to 24$\%$ of O being locked in dust grains. This is consistent with other studies, which find that O is present in substantial amounts in dust. \citet{Jenkins2009} estimates an excess of O in dust with respect to what is expected for the formation of known oxygen-rich compounds, in what he calls the `oxygen crisis'. We estimate the Si abundance into dust for maximum depletion as [Si/H]$_{\rm{dust}}$ = 40 $\pm$ 4\,ppm, assuming [Si/H]$_{\rm{ISM}}$ = 43 $\pm$ 4\,ppm from \citet{Zuo2021} and \citet{Hensley2021}. Therefore, up to 93$\%$ of Si is in the form of dust at maximum levels of depletion. The abundances in dust of O and Si are consistent with O being largely incorporated in silicates. For pure olivine
\ch{Mg_{2x}Fe_{2(1-x)}SiO_{4}}, 4[Si/H]$_{\rm{dust}}$ = 120\,ppm, consistent, within the uncertainties, with the O abundance in dust. Some oxygen may also be incorporated in iron oxides. Studies on dust composition, such as \citet{Roman-Duval2022a, Roman-Duval2022b, Konstantopoulou2024} find that O is the most abundant element by mass in dust in all the galactic environments studied (MW, LMC, SMC, QSO- and GRB-DLAs). Their results on dust composition suggest the presence of significant amounts of pyroxenes (\ch{MgSiO3}) in dust, as observed also by \citet{Mattsson2014}. Specifically, \citet{Konstantopoulou2024} estimate the maximum amount of O in dust to be 181\,ppm and a minimum 154\,ppm in the MW, which is consistent with our findings. Nevertheless, with [O/H]$_{\rm{dust}}$ = 141 $\pm$ 60\,ppm (as derived here) and typical gas-phase interstellar abundance of [O/H]$_{\rm{gas}}$ = 310\,ppm \citep{Meyer1998}, we derive a total O abundance of 450\,ppm. This is considerably lower than the interstellar abundance of [O/H]$_{\rm{ISM}}$ = 589 $\pm$ 68\,ppm, implying 138\,ppm (= 589\,ppm - 451\,ppm) of O remains unaccounted for in the ISM, consistent with earlier studies of \citet{Whittet2010} and \citet{Wang2015}. Therefore, the missing O problem is not solved, but the tension is not strong (1.5$\sigma$ level).

[O/Ti] is our most sensitive dust tracer due to the large difference in the refractory properties of O and Ti ($\Delta$B2$_{\rm{OTi}}$ = 1.44) and it correlates well with both [Zn/Fe] (see Fig. B.1, available on Zenodo) and with [Si/Ti] (see Fig. \ref{fig:siti_corr}). The strong correlation between the relative abundances [O/Ti] and [Si/Ti] is partially by construction, due to the fact that Ti is present in both x- and y-axis. The steeper slope than the 1:1 relation between [O/Ti] and [Si/Ti] of Fig. \ref{fig:siti_corr} indicates that O is depleted into dust much less severely than Si, which agrees with the results from \citet{DeCia2016, Roman-Duval2021, Konstantopoulou2022}. O is a volatile element, that remains for longer time in the gas-phase than Si, and has a weaker tendency to incorporate into dust compared to Si which is more refractory.

\subsection{Sulfur depletion in the diffuse ISM}
\label{sec: sulfur}

In the past, S was believed to not deplete into dust grains in the diffuse ISM, until the study of \citet{Jenkins2009}. Later studies on dust depletion of metals in the diffuse ISM confirmed that S, being a volatile element, is mildly depleted into dust grains \citep{DeCia2016, Roman-Duval2021, Konstantopoulou2022}. The panels showing [S/Ti] and [S/Mg] vs. [Si/Ti] in Fig. \ref{fig:siti_corr} give us insights into the dust depletion behavior of S in the diffuse ISM. The strong correlations between the relative abundances of metals show they follow the same sequences. The strong correlation between [S/Ti] and [S/Mg] with [Si/Ti] shown in Fig. \ref{fig:siti_corr} suggest that S is indeed depleted into dust grains at a lesser degree than Ti and Mg. If S would not deplete into dust grains one would expect a flat trend between [S/Ti] and [S/Mg] vs. [Si/Ti], which is clearly not observed. We note that the panel showing [S/Ti] vs. [Si/Ti] involves Ti in both x- and y-axis. This means that a strong correlation is created by construction. The steep slope between [S/Ti] and [Si/Ti] indicates that S is indeed depleted into dust but at a lesser degree than Si, and both S, Si deplete less heavily than Ti. These are consistent with the findings of \citet{DeCia2016, Roman-Duval2021, Konstantopoulou2022}. Other recent works, such as \citet{Yang2024}, discuss the depletion of S in the ISM, and specifically examine the possibility of the incorporation of S into dust grains or molecular forms that are less easily detected. This study reinforces the result that S is depleted, even in the diffuse ISM, in agreement with previous studies.

\subsection{Variations of the dust depletion with interstellar conditions}
\label{sec: molecular_fraction}

\begin{figure}[!ht]
    \centering
    \includegraphics[width=\columnwidth]{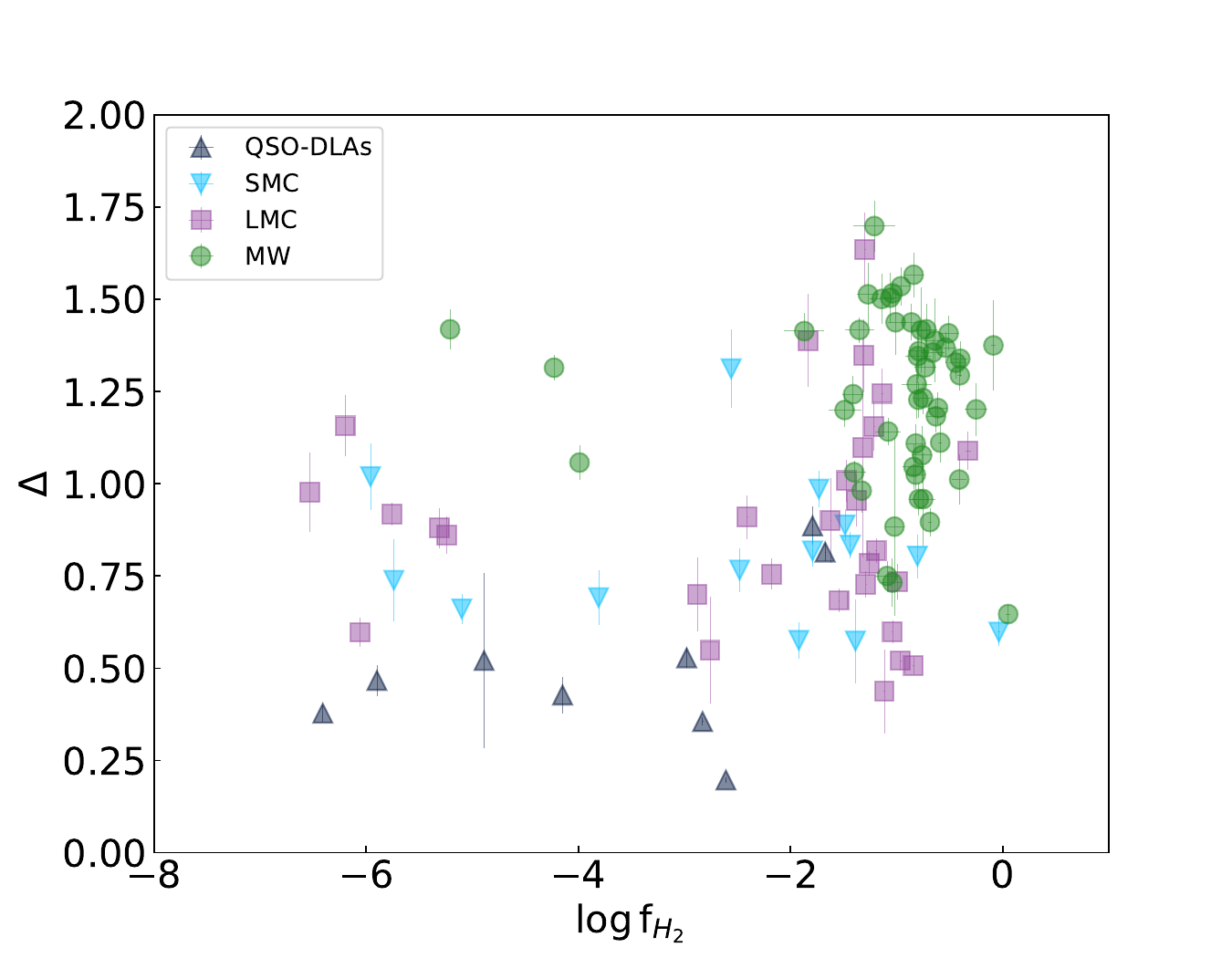}
    \caption{Overall strength of depletion $\Delta$ with respect to the molecular fraction $f_{H_{2}}$. The black triangles are for QSO-DLAs, the blue triangles for the SMC, the purple squares for the LMC and green circles for the MW.}
    \label{fig:fh2_Delta}
\end{figure}

The formation of molecules often occurs on the surface of dust grains. Therefore, we investigate the variation of the overall strength of depletion with the molecular fraction, defined as $f_{H_{2}}$ = $2\rm{N}(H_{2})/[\rm{N}(\ion{H}{I}) + \rm{N}(H_{2})]$, which expresses the fraction of hydrogen nuclei bound to H$_{2}$. Figure \ref{fig:fh2_Delta} shows the relation between the overall strength of depletion with respect to the molecular fraction $f_{H_{2}}$ for the MW, LMC, SMC and QSO-DLAs. We find no significant trend with the dust depletion $\Delta$. This is consistent with the findings of \citet{DeCia2024}, who find no clear correlation between the molecular fraction and the amount of dust depletion, which they trace with [Zn/Fe] (see their Fig. 12), and with \citet{Tumlinson2002} who find no clear correlation between $f_{H_{2}}$ and the color excess $E(B-V)$ (see their Fig. 7). Dust depletion probes the warm neutral medium of the ISM. On the other hand, $H_{2}$ molecules are known to form in the molecular medium, it is likely that they do not survive or form as abundantly in the warmer and diffuse phase of the ISM.

There is no clear variation in the abundances [O/H], [C/H] with interstellar conditions, for example with the molecular fraction $f_{H_{2}}$. This is consistent with previous results from \citet{Cardelli1996}, who found the gas-phase [C/H] abundance to be about 140\,ppm, independent of the interstellar conditions, such as $f_{H_{2}}$ (see their Fig. 2). It is also consistent with \citet{Meyer1998}, who found the gas-phase [O/H] abundance to be about 310\,ppm, also independent of the interstellar conditions. The agreement between the different studies is reassuring, since such gas-phase abundances of [C/H] and [O/H] have been frequently used to constrain dust models, such as \citet{Mathis1996, Li1997, Jones2013, Zuo2021, Zuo2021b}.

Additionally, we do not expect to have variations of the dust depletion with the hydrogen number density n$_{H}$ of the dense molecular clouds. The [O/H] and [C/H] abundances and the estimated overall strength of depletion $\Delta$ do not vary with the n$_{H}$ measurements from \citet{Meyer1998} and \citet{Klimenko2024}, in agreement with the conclusions of \citet{Meyer1998}. We note that \citet{Jenkins2009} found a relation between the strength of depletion (F$_{\ast}$) and the average n$_{H}$ in the warm neutral medium and cold neutral medium.

\section{Summary and conclusions}
\label{sec: conclusions}

Each metal exhibits a different tendency to deplete into dust grains, thus, their relative abundances can allow us to trace the amount of dust. In this work, we present a novel method to characterize the depletion of metals into dust using known correlations among relative abundances of metals that qualify as tracers of dust. This new method aims at taking full advantage of the strengths of the two main existing methods of estimating dust depletion (the F$_{\ast}$ and relative methods), while overcoming their weaknesses.

Using a comprehensive literature sample, we compiled the column densities of 17 metals for diverse galactic environments, including the MW, LMC, SMC, and QSO- and GRB-DLAs. To avoid complications arising from nucleosynthesis effects, we divided our elements into two groups: group 1, comprised of the $\alpha$-elements (O,
S, Si, Mg, and Ti), and group 2, comprised of the non-$\alpha$ elements (Fe, Cr, Ni, P, Kr, Ge, Co, Cu, Al, C, Cl, and Zn). We observed the linear correlations among all valid observed relative abundances of metals [X/Y] (dust tracers) across all the studied environments. 

Our method uses all the available dust tracers and the resulting parameters describing their correlations to estimate the overall strength of depletion, $\Delta$, along individual lines of sight. All dust tracers are correlated with each other, so there is no preferential use of any of them in the determination of the overall strength of depletion. We de-projected these correlations with respect to one reference dust tracer, [Si/Ti], so that the value of $\Delta$ is comparable to the observed values of [Si/Ti]. In cases where [Si/Ti] was not observed, we used the correlations between [Si/Ti] with [Ge/Ni] or [Zn/Fe] observed for the whole sample.

Moreover, we provide updated dust depletion estimates, on Cl, P, Ni, Cu, Cr, Zn, Ge, Fe, C, and Kr, which  directly result from our methodology. To benefit users, we used MW data to also provide dust depletion estimates on O, Si, Mg, and Ti, which result from the relative abundances of a mix of $\alpha$-elements with Kr (a refractory element). For the development of our methodology and the derivation of the overall strength of depletion, we strictly avoid mixing $\alpha$-elements with refractory elements when defining the relative abundances [X/Y].

In addition, we provide users with straightforward tools to determine the overall strength of depletion along individual lines of sight ($\Delta$), the gas-phase metallicity ([$\rm{M/H}$]), and any deviations due to a nucleosynthesis effect, from the study of the abundance patterns, when H is known. When H is unknown, we guide users on using metal patterns to determine the dust depletion and any deviations due to nucleosynthesis effect. Lastly, we offer simple guidelines on using any observed relative abundances of metals [X/Y] (or relative metal patterns) to estimate the overall strength of depletion. For these applications, we provide the required coefficients B$_{X}$ and C$_{XY}$. This novel method offers valuable insights into the chemical evolution of galaxies and the role of dust in shaping their galactic environments.

Data availability: Table 2 and appendices B-C are available on Zenodo at \url{https://doi.org/10.5281/zenodo.13889072}.

\begin{acknowledgements}
We thank the anonymous referee for the useful and constructive comments that improved this manuscript. Edward B. Jenkins passed away just before the submission of this paper. Ed was one of the great fathers of UV astronomy and his scientific work on the ISM was immensely impactful and inspiring. We are grateful to have had the opportunity to work with and learn from him. His legacy will live on.
CK, ADC, TRH and AV acknowledge support by the Swiss National Science Foundation under grant 185692. This research has made use of NASA’s Astrophysics Data System. 
\end{acknowledgements}

\bibliographystyle{aa}
\bibliography{ref}

\clearpage

\appendix

\section{Testing our methodology without Zn}
\label{excl_zn}

When defining the two groups we included Zn in the non $\alpha$-element group. The nucleosynthetic origin of Zn is complex and it is not considered an $\alpha$-element nor an Fe-peak element. However, prior studies \citep[e.g.,][]{Saito2009} have shown that Zn generally traces Fe in the metallicity range that we are studying here (-2 $\leq$ [M/H] $\leq$ 0). Additionally, [Zn/Fe] is the strongest dust tracer in this study, demonstrated by the large difference between the refractory indices of Zn and Fe ($\Delta$B2$_{{ZnFe}}$ = 0.99) and the large amount of data points. Therefore, Zn is included in group 2. 

However, we acknowledge that Zn can show a small enhancement, but of smaller amplitude than the $\alpha$-elements \citep[e.g.,][]{Barbuy2015, Velichko2024}. This small amplitude is, however, not affecting our analysis. First, the correlations involving Zn are not affected, as shown in Fig. \ref{fig:siti_corr}, where the QSO-DLA points (black triangles) pass through zero, despite the scatter caused by inhomogeneities in the \citet{Berg2015} sample. Second, we tested the analysis by excluding Zn and our results remain unaffected.

Although there is still some debate on the reliability of [Zn/Fe] as a dust tracer, it remains the strongest dust tracer in our study, demonstrated by the large difference between the refractory indices of Zn and Fe ($\Delta$B2$_{{ZnFe}}$ = 0.99). Therefore, [Zn/Fe] is used in our methodology as an indirect reference dust tracer and Zn is involved in several other dust tracers, as well, e.g., [Zn/Ni], [Zn/Cr], [Zn/Co]. To assess the impact of Zn on our results, we conduct a comparative analysis by recalculating the overall strength of depletion, $\Delta$, without including Zn. For this test, we limit the analysis to a subset of lines of sight with available Si and Ti measurements, using only [Si/Ti] as a reference dust tracer.

\begin{figure}[!h]
    \centering
    \includegraphics[width=\columnwidth]{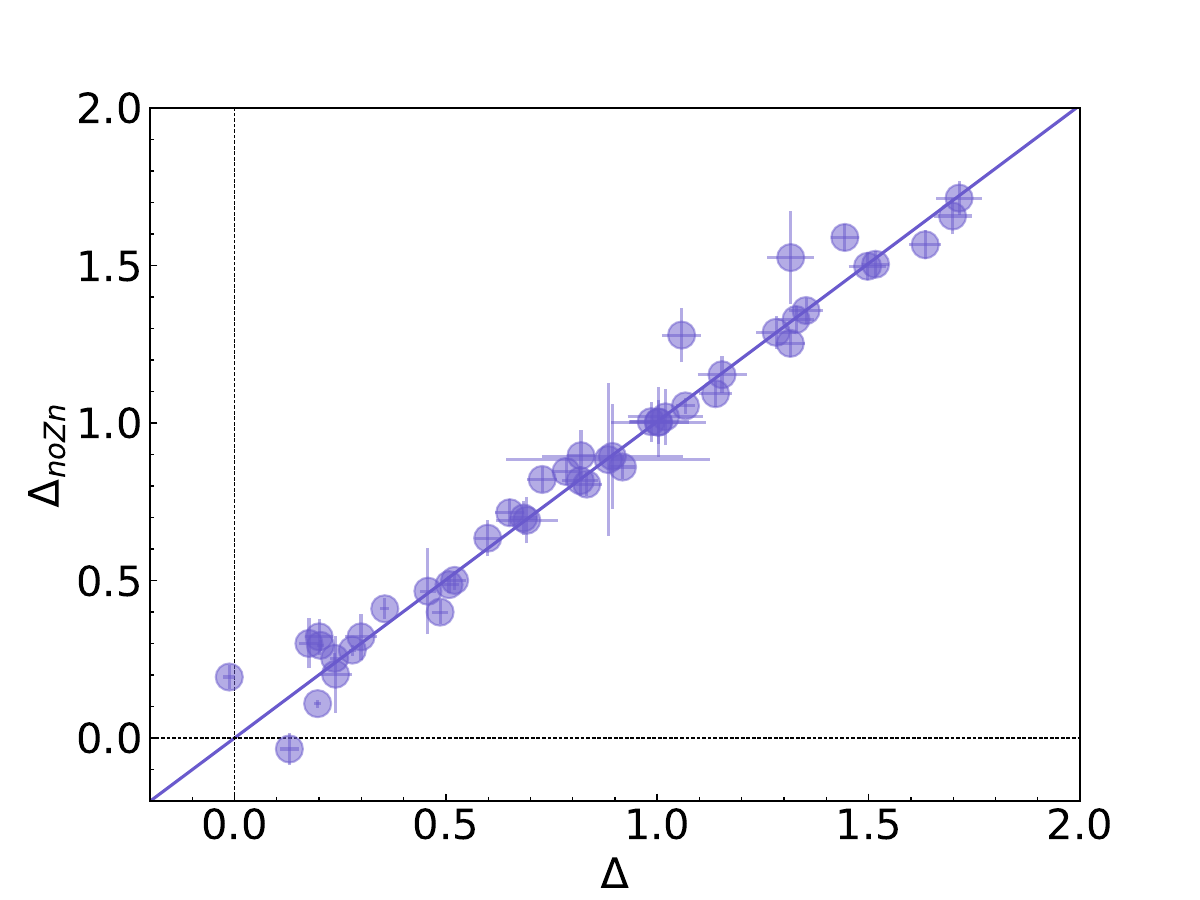}
    \caption{Comparison of the overall strength of depletion obtained from Eq. \ref{eq:Delta} including Zn ($\Delta$) and the overall strength of depletion after excluding Zn from the analysis ($\Delta_{noZn}$).}
    \label{fig:comparison_delta}
\end{figure}

The comparison between the overall strength of depletion calculated with ($\Delta$) and without Zn ($\Delta_{noZn}$) is shown in Fig. \ref{fig:comparison_delta}. The data follow a 1:1 relationship, represented by the slope (B$_{X}$ = 1.00 $\pm$ 0.02) of the linear fit. This reassures the validity of our methodology. Additionally, [Zn/Fe] shows strong correlations with multiple other dust tracers, as illustrated in Figure B.1. This analysis confirms that the inclusion of Zn does not compromise the reliability of our method, and the use of [Zn/Fe] as an indirect reference dust tracer is justified. This comparative analysis supports the non-privileged treatment of any dust tracer in our methodology, since all dust tracers are correlated with each other in a multidimensional space.

\end{document}